\documentclass[12pt]{elsarticle}
\usepackage{amssymb}
\usepackage{amsmath,epsfig}
\usepackage[latin1]{inputenc}
\usepackage{graphicx}
\usepackage[english]{babel}

\usepackage{xspace}
\usepackage{subfigure}
\usepackage{float}
\usepackage{multirow}
\usepackage{hyperref}
\usepackage{placeins}
\usepackage[percent]{overpic}
\usepackage{eurosym}
\usepackage{tablefootnote}
\usepackage{hyperref}
\usepackage{array}
\newcolumntype{L}[1]{>{\raggedright\let\newline\\\arraybackslash\hspace{0pt}}m{#1}}
\newcolumntype{C}[1]{>{\centering\let\newline\\\arraybackslash\hspace{0pt}}m{#1}}
\newcolumntype{R}[1]{>{\raggedleft\let\newline\\\arraybackslash\hspace{0pt}}m{#1}}

\usepackage{lineno}

\journal{Nuclear Instruments and Methods in Physics Research, }

\begin{document}

\begin{frontmatter}

\title{Measurements of timing resolution of ultra-fast silicon detectors with the SAMPIC waveform digitizer 
}
\author[add1]{D. Breton}
\author[add3]{V. De Cacqueray\fnref{fn1}}
\author[add3]{E. Delagnes}
\author[add4]{H. Grabas}
\author[add1]{J. Maalmi}
\author[add5]{N.~~Minafra\fnref{fn2}}
\author[add6]{C. Royon}
\author[add3]{M. Saimpert\corref{cor1}}
\ead{matthias.saimpert@cern.ch}

\address[add1]{CNRS/IN2P3/LAL Orsay, Universit\'{e} Paris-Saclay, F-91898 Orsay, France}
\address[add3]{IRFU, CEA, Universit\'{e} Paris-Saclay, F-91191 Gif-sur-Yvette, France}
\address[add4]{Santa Cruz Institute for Particle Physics UC Santa Cruz, CA 95064, USA}
\address[add5]{Dipartimento Interateneo di Fisica di Bari, Bari, Italy, CERN, Geneva, Switzerland}
\address[add6]{University of Kansas, Lawrence, USA}
\cortext[cor1]{Corresponding author.}
\fntext[fn1]{now at: \textit{LSCE, CEA, Universit\'{e} Paris-Saclay, F-91191 Gif-sur-Yvette, France.}}
\fntext[fn2]{now at: \textit{University of Kansas, Lawrence, USA.}}

\begin{abstract}
The SAMpler for PICosecond time (\textsc{SAMPIC}) chip has been designed by a collaboration including CEA/IRFU/SEDI, Saclay and CNRS/LAL/SERDI, Orsay. It benefits from both the quick response of a time to digital converter (TDC) and the versatility of a waveform digitizer to perform accurate timing measurements. 
Thanks to the sampled signals, smart algorithms making best use of the pulse shape can be used to improve time resolution.
A software framework has been developed to analyse the \textsc{SAMPIC} output data and extract timing information by using either a constant fraction discriminator or a fast cross-correlation algorithm.
SAMPIC timing capabilities together with the software framework have been tested using pulses generated by a signal generator or  by a silicon detector illuminated by a pulsed infrared laser.
Under these ideal experimental conditions, the \textsc{SAMPIC} chip has proven to be capable of timing resolutions down to 4~ps with synthesized signals and 40~ps with silicon detector signals.
\end{abstract}


\end{frontmatter}


\section{Introduction}
\label{Introduction}

At the Large Hadron Collider (LHC) at CERN~\cite{Evans:2008zzb}, which is the highest energy proton-proton
collider in the world with a designed center-of-mass energy of 14~TeV, special 
classes of events can be studied where protons are found to be intact after collisions. These
events are called ``diffractive" in the case of gluon exchanges. They can also originate from
photon exchanges as well and then called photon-induced processes. The physics motivation for their study is a better understanding
of diffraction in terms of QCD~\cite{yellow_report,qcd1,qcd2} and the search for physics beyond the standard model~\cite{yellow_report,anomalous0,anomalous00,anomalous1,anomalous2,anomalous3,anomalous4}. 

The intact protons scattered at small angles
can be measured using dedicated detectors hosted in roman pots, a movable section of the vacuum chamber that can be inserted a few milimeters away from 
the beam at more than 200~m from each side of the main central ATLAS~\cite{1748-0221-3-08-S08003} or CMS~\cite{Chatrchyan:2008aa} detectors. 
In order to measure rare events at the LHC, the
luminosity (\textit{i.e.}~the number of interactions per second) has to be
as large as possible. In order to achieve this goal, the number of interactions
per bunch crossing is planned to be very large, up to 40-70 during the second LHC run (2015-2018). The projects aiming to measure intact protons at high
luminosity in the ATLAS and CMS/TOTEM~\cite{Anelli:2008zza} experiments are called respectively
AFP (ATLAS Forward Proton detector)~\cite{Soni:2010du} and CT-PPS (CMS/TOTEM-Precision Proton
Spectrometer)~\cite{Albrow:1753795,afpctpps}.

In this context, timing measurements are crucial in order to determine if the intact 
protons originate from the main
hard interaction or from additional interactions in the same bunch crossing, called pileup interactions in the following. 
Indeed, if two intact protons are detected in coincidence on each side of the main interaction point, their time-of-arrival can be used to reconstruct their vertex. The latter can then be compared to the vertex reconstructed from the high energy particles detected in the central detector and one can estimate the compatibility between the two. Measuring the arrival time of protons with a typical precision of 10 ps RMS\footnote{Root Mean Square (RMS). It indicates that the precision is evaluated from the spread of the whole distribution of the measurements and therefore do not correspond to a standard deviation $\sigma$ returned by a Gaussian fit for instance.} allows to discriminate between hard interaction vertices and pileup interactions up to a precision of about $\pm$ 2 mm. For 40 interactions occurring in the same bunch crossing at the LHC, such a
precision leads to a background reduction by a factor close to 40~\cite{matthias}.

Timing measurements have also many other applications, for example in
medical imaging. Indeed, \textsc{PET} imaging would highly benefit from a 10~ps timing precision~\cite{10.3389/fonc.2016.00009}: a large fraction of 
fake coincidences would be automatically suppressed thanks to the reduced time window, the emission vertex of the two back-to-back photons being point-like despite the possible large extension of the tumours ($\simeq$~10~cm for a full body examination). That would increase the Signal over Noise Ratio (SNR) and so decrease the amount of recorded data required for an examination. Smaller tumours may be spotted as well. Moreover, depending on the SNR improvement, additional filters applied typically during the offline image reconstruction to improve image quality could be no longer necessary and real time image formation could be performed. Those improvements would allow to reduce the exposure of the patient to the radioactive tracer and to repeat immediately the tomography if needed.

The \textsc{SAMPIC} chip~\cite{sampic_paper} has been designed to achieve about 5 ps RMS timing precision. In this paper, the second SAMPIC release is used (called V1). We present the results of various tests carried out using pulses synthesized by a signal generator and silicon detectors illuminated by a pulsed infrared laser. The analysis is performed with a dedicated software framework developed to analyse \textsc{SAMPIC} data. 

In Section~\ref{sampic}, we give a brief description of the SAMPIC chip, stressing the
advantages with respect to previous technologies. Section~\ref{online} is dedicated 
to the hardware and the online software used to acquire data. The offline software employed for timing reconstruction is presented in Section~\ref{Software} while the tests 
of the SAMPIC chip are presented in Section~\ref{Tests}, first with synthetic signals and then using fast silicon detectors. Finally, a summary of the SAMPIC performance measured in this paper is reported in Section~\ref{conclusion}.

\section{The \textsc{SAMPIC} chip}
\label{sampic}
\subsection{TDCs}

Current systems usually rely on Time to Digital Converters (TDC) to measure precise timing. A TDC uses a counter that provides coarse timing associated with a Delay Locked Loop (DLL) doing a finer interpolation of the latter. The timing resolution is then often limited by the
DLL step and with most advanced Application Specific Integrated Circuits 
(ASIC), one can achieve a resolution of about 20 ps RMS~\cite{NewTDC_paper}.\footnote{new developments at CERN target 5~ps~RMS.}

An additional drawback of this technique for analog pulses (such as the ones produced in a typical particle detector) is that the TDC relies on a digital input signal. Therefore, one needs first to convert the signals prior to any measurement. 
This conversion is typically done using a fast discriminator. However, the amplitude dispersion of the signal may induce time walk effects on the digital output signal, even if they are usually corrected via time over threshold measurements.
Furthermore, the discriminator limits the precision of the measurement and therefore introduces an additional jitter to the system. The timing resolution of such a system is then given by the quadratic sum of the TDC and the discriminator resolution. 

\subsection{Waveform TDC}

To overcome this limitation, a new approach has been proposed, based on the innovative principle of a waveform-based TDC (WTDC). In such a system, a fast digitizer based on an analog memory is added in parallel with the delay line. It is used to acquire the part of the input analog signal which is relevant for timing. Digital processing such as interpolation applied on the digitized data allows to reach timing resolution of few ps, far better than the time step of the DLL. 
A discriminator is anyhow present, but only used to assert a trigger signal and therefore not anymore in the critical timing path.

\begin{figure}[htpb]
\centerline{%
\includegraphics[width=1.0\textwidth]{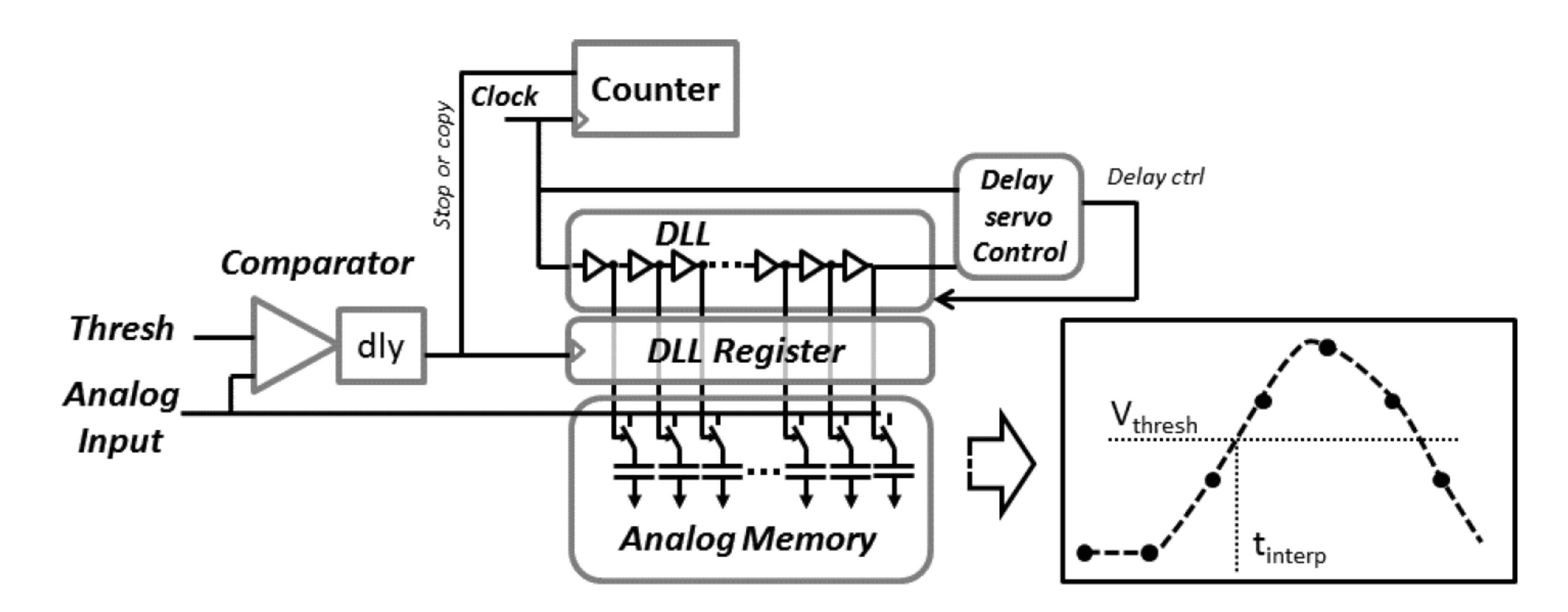}}
\caption{Schematic of one channel of the SAMPIC WTDC.}
\label{Fig2}
\end{figure}

\textsc{SAMPIC} implements the WTDC approach in an ASIC for 16 independent input channels. A basic schematic of one SAMPIC channel is shown on the Figure~\ref{Fig2}. The main features of the SAMPIC V1 are summarized in the Table~\ref{Figa}. This version is only a slight evolution of the previous SAMPIC V0 chip, allowing for a better linearity. Detailed information about SAMPIC V0 can be found in~\cite{sampic_paper}.

In Section~\ref{online}, the integration of \textsc{SAMPIC} is described.

\begin{table}[htbp]
\centering
\small
\begin{tabular}{l l l }
\hline\hline
  & Performance measured & Comment\\
\hline
 & & \\
Technology  & AMS CMOS $0.18~\mu m$ & \\
Number of channels  & 16 & \\
Power consumption  & 180~mW (1.8~V supply) & \\
Capacitor array depth & 64 & \\
Discriminator noise  & 2~mV & \\
Sampling speed & 1.6 - 8.4 GS/s & Up to 10.2~GS/s on \\
                &                 & the 8 first channels. \\
3dB Input bandwidth   & 1.6 GHz & \\
ADC precision  &  8 to 11 bits & Trade-off time vs\\
               &               & resolution. \\
Noise &  $< 1.3$ mV RMS & \\
Input dynamic range  &  1~V (0.1~V to 1.1~V) & \\
Conversion time  &  200 ns for 8~bits - 1.6 $\mu s$ on 11 bits & \\
Readout time   &  37.5~ns + 6.25~ns/sample at 160~MHz & A readout up to  \\
               &                                      &  200~MHz is possible. \\
Timing precision:  &   & \\
\textit{before calibration} & 15~ps RMS & \\
\textit{after calibration}   &  $< 5$ ps RMS & \\[1ex]
\hline
\end{tabular}
\caption{Table of SAMPIC V1 main performance. }
\label{Figa}
\end{table}

\section{Acquisition board and control software}
\label{online}

The SAMPIC chip is integrated in an \emph{out-of-the-box} configuration that includes one or two mezzanine boards embedding the chip plugged on a motherboard permitting its control and readout. Software in charge of controlling the acquisition, displaying and saving data is also available. The software also allows to perform real-time measurements on the signals.

\begin{figure}[htb]
\centerline{%
\includegraphics[width=0.5\textwidth]{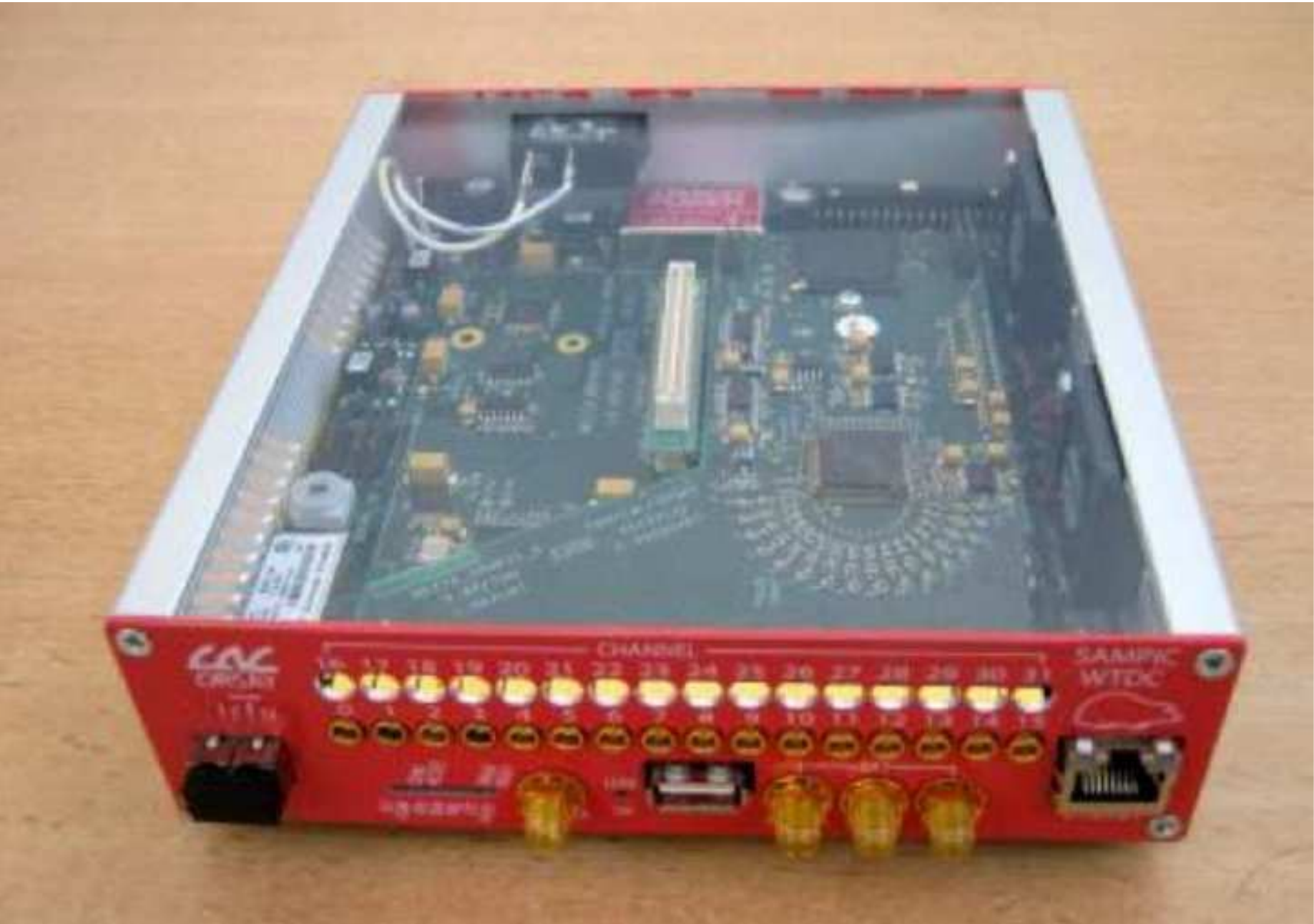}}
\caption{Picture of a 16-channel SAMPIC device.}
\label{Fig_acquisition}
\end{figure}

The full device, shown in Figure~\ref{Fig_acquisition}, requires a simple 5 V power supply and a PC that can be connected through USB 2.0, Gigabit Ethernet or optical fibre. Moreover, it is possible to supply an external clock and some control signals, for instance to setup an external trigger or a signal that vetoes the trigger.
The device has 16 or 32 MCX connectors for the analog input, depending on the number of SAMPIC chips used (1 or 2).

The control software is available for Windows platforms and based on LabWindows libraries. A multi-platform version is under development.
The main parameters that can be set using the software are:
\begin{description}
	\item[Sampling parameters:] sampling frequency (from 1.6 to 10.2 GS/s), baseline (from 0.1 to 1.1 V), Analog to Digital Conversion (ADC) range (8, 9, 10 or 11 bits).
    \item[Trigger parameters:] trigger mode, thresholds and polarity of the signal.
    \item[Acquisition parameters:] data format (ASCII or binary) and maximum file size, length of the acquisition (time or number of hits).
\end{description}

Moreover, it is possible to correct the acquired data through the control software~\cite{sampic_paper}. Two types of calibration are implemented: the first one improves the ADC linearity of each cell and the second one corrects the non uniformity of the sampling period in the analog memory, which is related to the chip architecture and geometry.
The calibration procedure needs to be repeated for each set of acquisition parameters (\textit{i.e.} sampling frequency and number of ADC bits).
The software takes care of storing the configuration parameters for all the different acquisition scenarios and makes use of them to correct the acquired data.
It should be noted that without time calibrations, the SAMPIC chip is still able to achieve a very precise time resolution, of the order of 15~ps RMS at 6.4 GS/s~\cite{sampic_paper}. 

Finally, the software can also be used to display acquired data and to perform some preliminary analysis. For instance, it is possible to measure the fluctuations on each channel when no signal occurs (mean, RMS) and to compute online the time difference between two channels via basic algorithms without any pre-processing.

\section{Offline software}
\label{Software}

\subsection{Data format definition}
\label{DataDef}

The data acquired using the SAMPIC chip are processed by the acquisition board and can be saved in a binary or ASCII file using the control software.
A preliminary estimation of the time differences can be made online using the latter. However, an offline analysis is very useful during the development stage in order to compare easily several time algorithms and implement some data cleaning and/or pre-processing if required.

The offline software is a C++ code based on the \textsc{ROOT} framework.\footnote{\url{https://root.cern.ch/}}
First, the sampled waveforms are extracted from the data acquired with the online software and stored in a hierarchical structure (TTree). Data are organized according to their timestamps. In particular, if an external trigger is provided, two or more hits are considered belonging to the same ``event'' when the differences between the channel and the trigger timestamps are smaller than $\Delta t_{max}$, after being corrected for a fixed trigger latency.
Otherwise, all coincident hits, \textit{i.e.} with a time difference between their respective starting time less than $\Delta t_{max}$, are assigned to the same event. The value of $\Delta t_{max}$ needs to be set by the user.

For each acquisition, the baseline, \textit{i.e.} the average sampled voltage when no event occurs, is measured from the first $n_b$ recorded points, where $n_b$ must be specified by the user. It is then recorded in the data tree. The first $n_b$ recorded points are also used to compute for each channel the RMS noise, which is defined as the RMS of the average sampled voltage when no event occurs. Other relevant information such as sampling period, channel number and ADC counts are also saved. This intermediate step is useful to disentangle the data acquisition from the analysis part, foreseeing the usage of the same analysis code for future version of the hardware and with other sampling devices such as an oscilloscope.



\subsection{Time measurement analysis}
\label{sec:time}

Several algorithms have been proposed to extract the timing of a digitized signal \cite{Bousselham2007, Bertuccio1992,Joly2010,Leroux2009, Genat:2008vc,Bardelli2004,grabas_thesis,Stricker2014,FNCC_lewis} .
Two time reconstruction algorithms have been selected among them and implemented in the offline software because of their known performance, their simplicity and their limited resources requirement for a potential hardware implementation. They can be run on any SAMPIC data tree via the offline software:

\begin{itemize}
\item Constant Fraction Discriminator (CFD)~\cite{Genat:2008vc,Bardelli2004,grabas_thesis}.
\item Cross-Correlation (CC)~\cite{grabas_thesis,Stricker2014,FNCC_lewis}.
\end{itemize}


For precise timing, the most common approach to compute the time of arrival of a sampled signal is to use a CFD algorithm~\cite{Genat:2008vc,Bardelli2004,grabas_thesis}. The arrival time is then defined as the instant when the signal crosses a threshold corresponding to a given fraction $0<R<1$ of the signal amplitude, set by the user beforehand.
This threshold definition makes the results almost independent from the signal amplitude (suppression of the time walk effect). The signal amplitude is retrieved from a parabola interpolation (based on three points) and the threshold crossing value is obtained by using a linear interpolation of the sampled points (based on 2 points). 

The time difference between two channels is then given by:

\begin{equation}
    \label {dt_cfd}
    \Delta t^{CFD} = (t_{0,2} + t_2^{CFD}) - (t_{0,1} + t_1^{CFD}),
  \end{equation}

\noindent where $t_1^{CFD}$, $t_2^{CFD}$ are the CFD threshold crossing times of channel 1 and 2. Those quantities are measured respectively from $t_{0,1}$, $t_{0,2}$, which correspond to the time of the first recorded point in channel 1 and 2.

Assuming that the time walk is corrected by an offline signal algorithm, the time resolution of an analog signal followed by a TDC is given by\footnote{The time resolution of an analog signal is given by the sum of 3 uncorrelated contributions: jitter, time walk and time drift. The time drift is due to slow processes and is usually negligible for fast timing, the time walk is assumed to be corrected and the jitter depends on the noise and the slope of the signal as shown in Equation~\ref{eq:cfd}. More details can be found in chapter 6 of~\cite{Minafra_PhD}. The jitter from the TCD is assumed to be uncorrelated.}:

\begin{linenomath}
\begin{equation}
\sigma_{\rm CFD}^2 = \left( \sigma_{\rm noise}\cdot\frac{\tau}{A} \right)^2 + \sigma_{\rm TDC}^2,
\label{eq:cfd}
\end{equation}
\end{linenomath}

\noindent where $\sigma_{\rm noise}$ is the RMS noise of the channel, $\frac{\tau}{A}$ is the rise time divided by the signal amplitude, assuming the signal to rise linearly, and $\sigma_{\rm TDC}$ represents the timing jitter of the TDC. 

This formula will be used in the following sections as a reference. In the case of a digital CFD, at least 2 samples are used to extract the timing. This improves the timing resolution ($\sigma_{\rm CFD}$) 
by a factor $\sqrt{3/2}$ with respect to the value given by Equation~\ref{eq:cfd}, see~\cite{eric_note} for more details. However, this improvement requires the noise and jitter of the samples to be uncorrelated, \textit{i.e.} if the noise and the jitter of the samples are fully correlated the reduction factor disappears. In the following, for comparison purpose, we decided to be conservative and to not consider the $\sqrt{3/2}$ potential improvement.


A refined version of the CFD algorithm has also been implemented in the software. It computes the average of the times returned by multiple CFDs with different ratios. In case more than two points are available along the rising edge, this technique often improves the resolution by using the timing information carried by more samples than the classic CFD implementation. The computation time for all CFD-based algorithms implemented in the offline software is of the order of 1~ms per event.\footnote{Computation performed with a 2.3 GHz processor including a 4 Gb memory.} 


The alternative method implements a cross correlation algorithm~\cite{grabas_thesis,Stricker2014,FNCC_lewis}. A correlation function is computed between the acquired signal and a template extracted from an independent set of acquisition with high statistics. The latter is usually built from a reduced time window around the rising edge of the signals. The maximum of the correlation function, obtained by varying the delay between the signal and the template, indicates the optimal superposition of the two.

Let $s$ be a signal sampled by SAMPIC and $n$ the number of points of the corresponding template $t$, defined in a sliding time sub-range of the SAMPIC acquisition window $[delay, delay+n-1]$. We recall that $s$ is defined in the whole SAMPIC acquisition window, which depends on the sampling frequency (see Section~\ref{sampic}). We start by considering the signal average in $[delay, delay+n-1]$:

\begin{equation}
    \label {mean}
    \overline{s}(delay)= \overline{s(delay), s(delay+1), ... , s(delay+n-1)}.
  \end{equation}

The correlation function is then defined as:

\begin{equation}
    \label {cc}
    C(delay)= \dfrac{\sum_{i=0}^{n-1} (t(i)-\overline{t}).(s(i+delay)-\overline{s}(delay))}{\sqrt{\sum_{i=0}^{n-1} (t(i)-\overline{t})^2 }. \sqrt{\sum_{j=0}^{n-1} (s(j+delay)-\overline{s}(delay))^2}}.
  \end{equation}
  
The $delay$ maximizing the correlation function $C$ for each signal can then be used to determine the time difference between two channels:

 \begin{equation}
    \label {dt}
    \Delta t^{CC} = (t_{0,2} + t_2^{CC}) - (t_{0,1} + t_1^{CC}),
  \end{equation}
  
\noindent where $t_1^{CC}$, $t_2^{CC}$ are the optimal delays obtained for channel 1, 2 and $t_{0,1}$, $t_{0,2}$ are the time of the first recorded points in the corresponding channels. 

The main steps performed by the cross correlation algorithm of the software can be summarized as follows:

\begin {itemize}

\item A template is generated for each channel from an independent, high statistics data sample acquired with the same experimental apparatus and under similar conditions as the data to be analysed. Baselines are subtracted and all signals are normalized to their amplitudes, interpolated at 1~ps and synchronized together using a CFD algorithm. The template is then computed by averaging all signal shapes in a reduced window defined around the rising edges of the synchronized signals.

\item For each hit to be analysed, the signal is interpolated to 1~ps and normalized to its amplitude. A coarse synchronization between the signal and the template is done using a standard CFD.

\item The maximum of the cross correlation function is then determined using 1~ps steps for the $delay$ variable (see Equation~\ref{cc}). In order to reduce the computation time, a limited $delay$ range around the coarse synchronization time $t_{synch.}$ discussed above must be chosen by the user. In the tests which will be presented in Section~\ref{Tests}, we use $[t_{synch.} - 250$~ps$, t_{synch.} + 250$~ps]. It has been checked that the maximum of the correlation function is never reached on the edges of the range.

\item The optimal $delay$ between the signal and the template is recorded for each signal and can be used to compute the time difference between two channels (see Equation~\ref{dt}). 

\end{itemize}


In principle, the cross correlation algorithm can exploit the information contained in all the sampled points of an acquisition whereas the CFD uses only a few points by definition, \textit{i.e.} the ones close to the maximum and the ones close to the threshold crossing.
Moreover, contrarily to the position of the CFD threshold, the position of the maximum of the correlation function is not biased by an inaccurate baseline subtraction of the signals 
and its value at maximum could in principle be used as a data quality estimator. 

However, the CC requires the generation of templates for the expected signals, which might not be straightforward for certain application with relatively high statistical fluctuations of the signal like particle detection with thin detectors.
It should be noted also that the computation time of the CC algorithm is significantly higher than for the CFD algorithm by about a factor 100. 

In conclusion, the offline software is able to measure the time difference between two signals via various signal algorithms (CFD, refined CFD, CC). The signal-template synchronizations via cross correlation are also recorded in the output. 

In Section~\ref{Tests}, results from experimental tests are reported. In those tests, the time difference between the signals is fixed and the measurement with \textsc{SAMPIC} is repeated many times under similar conditions. The RMS of the reconstructed time distribution is defined as the timing resolution of the acquisition chain and is used to measure its performance.






\section{Experimental measurements}
\label{Tests}

Different tests were undertaken to measure the \textsc{SAMPIC} performance and compare the results of the various algorithms implemented in the offline software. First, some preliminary tests done using signal synthesized by a signal generator are reported in Section~\ref{sec1}. Then, in Section~\ref{sec2}, the \textsc{SAMPIC} chip and offline software are tested together with ultra-fast silicon detectors (UFSDs)~\cite{HFWS, Cartiglia1} illuminated by a pulsed infrared laser. The UFSDs are new generation silicon detectors being developed at INFN, Torino, UC Santa Cruz and the Institute of Microelectronics of Barcelona for fast timing application. More details will be given in Section~\ref{sec2}.

\subsection {Tests with signal generator}
\label{sec1}

Preliminary tests of the \textsc{SAMPIC} chip are done using signals from a LeCroy pulse generator.\footnote{LeCroy 9214 -- 300 MHz.} Those signals are used to characterize the chip and to investigate the performance of the various algorithms under ideal conditions.
The generator is configured to produce pulses with a rise time smaller than $1 \, \rm{ns}$ and a width of 0.90 ns at a rate of 1 kHz. The amplitudes of the signals are reduced using several broadband attenuators (Wavetek, 3~GHz) and then split by a T-junction. Finally, the signals are sent to two different channels of SAMPIC through cables of different length in order to create a delay between them. 
The jitter between the two signals is assumed to be less than $ 1 \, \rm{ps}$ when they enter the chip.

The sampling frequency of \textsc{SAMPIC} is set to 6.4 GS/s for the entire test.
\textsc{SAMPIC} channels are in self-trigger mode with a threshold set according to the incoming signal amplitude. During the first series of tests, the delay between the two signals is fixed to about 5~ns while the signal amplitude is reduced progressively using the attenuators (\emph{amplitude test}). During the second series of tests, the amplitude of the generator signal is set to 1.2~V while the delay between the two signals acquired by the \textsc{SAMPIC} chip is varied from a few picoseconds to a few hundreds of nanoseconds (\emph{delay test}). 

\begin{figure}[t]
\centering
\includegraphics [width=0.7\textwidth]{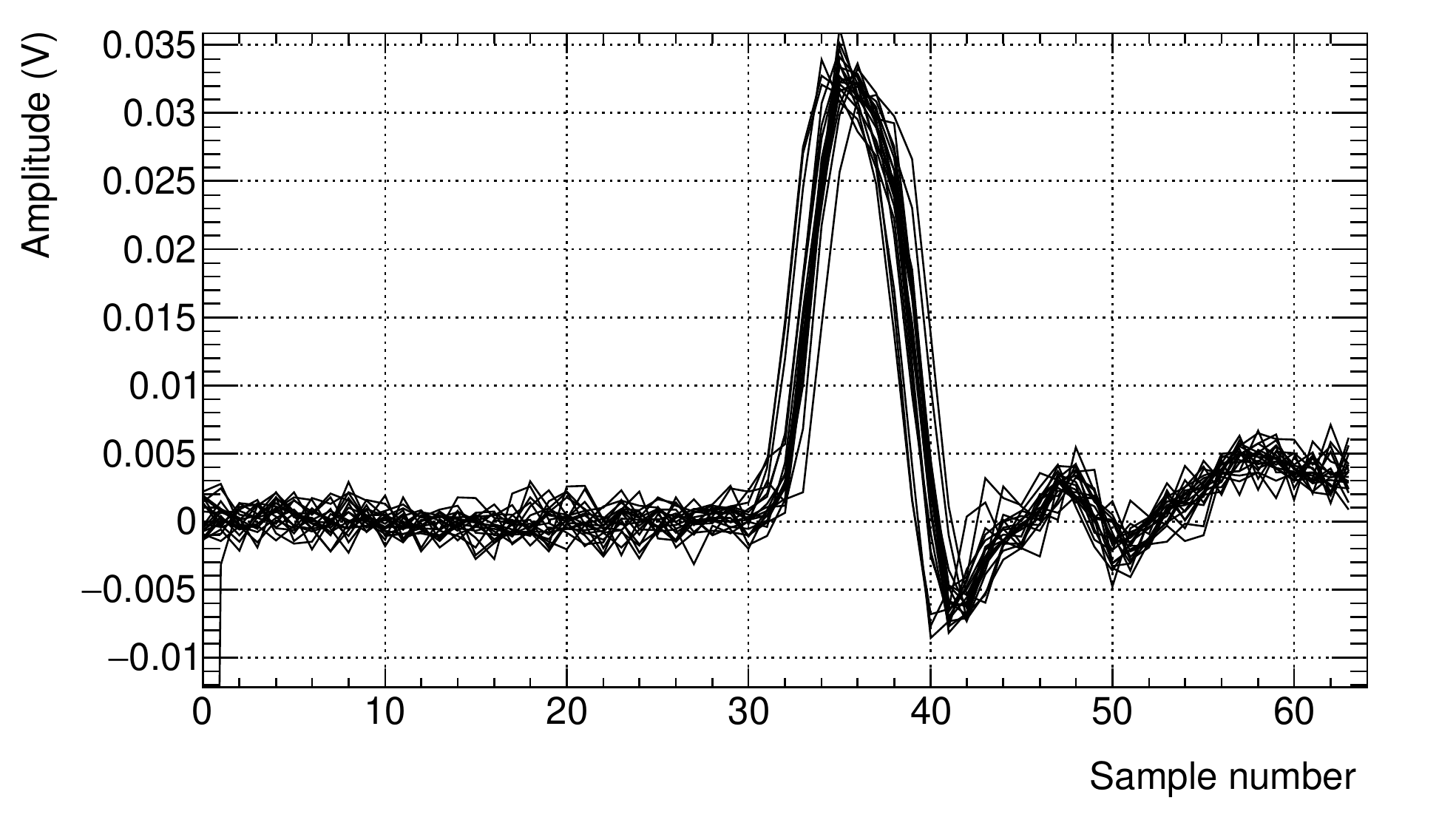}\\
\caption{Several signals with an amplitude of 33 mV acquired using the \textsc{SAMPIC} chip sampling at 6.4 GS/s are shown. The time spacing between two samples is about 156~ps. The signals are produced using a pulse generator and interpolated linearly for plotting purposes. The strong overlap of the various curves shows the low statistical fluctuations of the signal shapes. The first point is sometimes lost, which is a known feature of the chip related to the SAMPIC architecture.}
\label{elec_signals}
\end{figure}

Examples of signals acquired by one of the \textsc{SAMPIC} channels from 27~dB-reduced synthesized pulses ($\simeq$ 33~mV amplitude) are shown in Figure~\ref{elec_signals}. Some reflections of the pulse are arising from the impedance mismatch of the cable and the signal splitter. Due to the limited bandwidth of the cables and skin effect, we also noticed an attenuation of the signal at high delay ($>$~100 ns) by up to 30\% due to large cable length (delay test).

\begin{figure}[!htbp]
\centering
\includegraphics [width=0.7\textwidth]{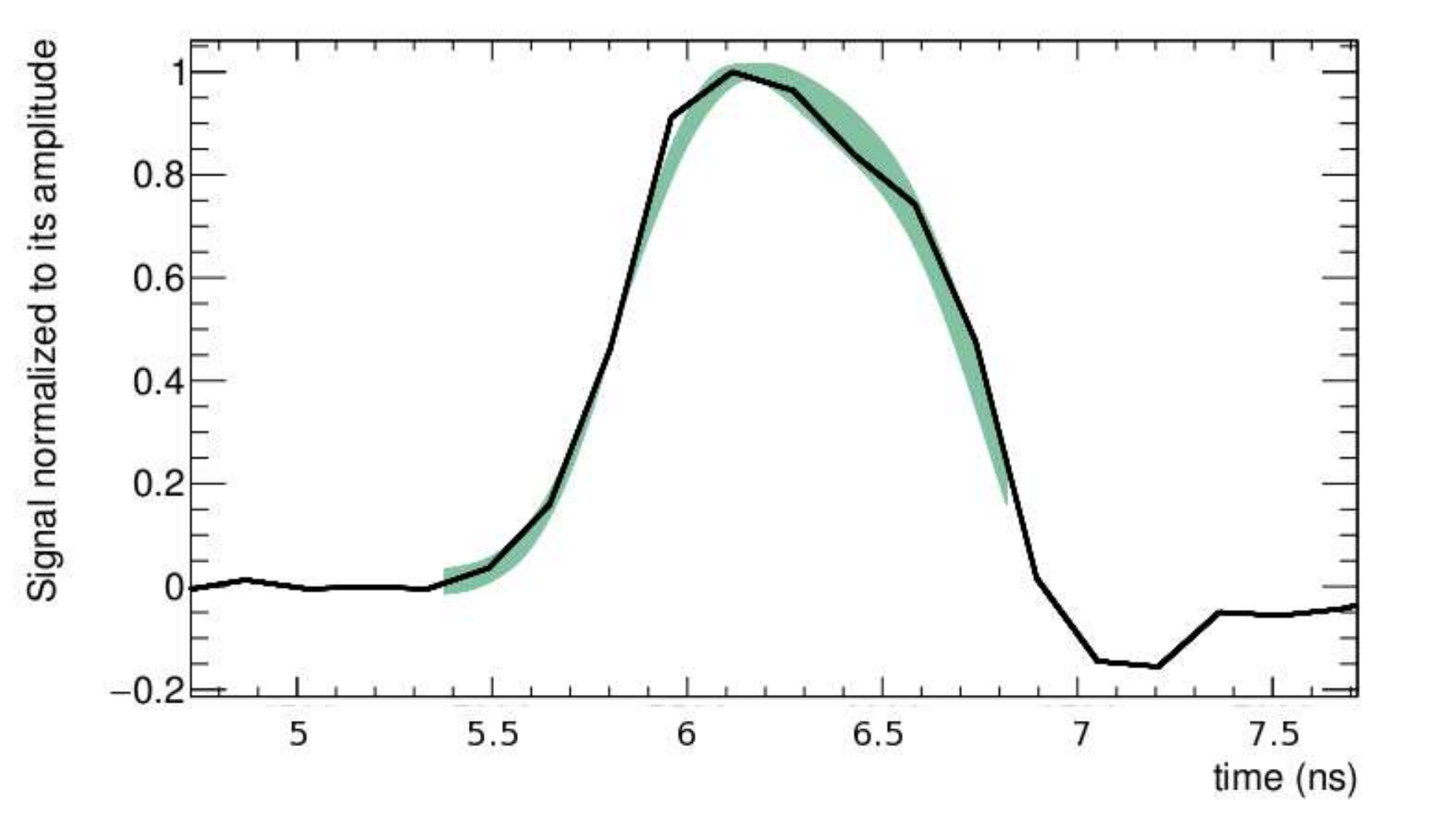}
\caption{Synchronization performed between a template (band) and a signal (line) by the cross correlation algorithm. The band of the template corresponds to its statistical dispersion (RMS). The signal corresponds to one of the sampling shown in Figure~\ref{elec_signals}.}
\label{sync_elec}
\end{figure}

For each data sample, 10,000 events are processed with the offline software (see Section \ref{Software}) using both the CFD and the cross correlation algorithm to reconstruct the time difference between the two channels.
The default CFD is configured to use a threshold at $R=0.5$ times the amplitude. A refined CFD algorithm averaging the results obtained with thresholds at $30\%, 35\%, \dots, 70\%$ of the amplitude is also performed. Furthermore, an example of synchronization between a template and a signal using the cross correlation algorithm is shown in Figure~\ref{sync_elec}.

\begin{figure}[t]
\centering
\includegraphics [width=0.732\textwidth]{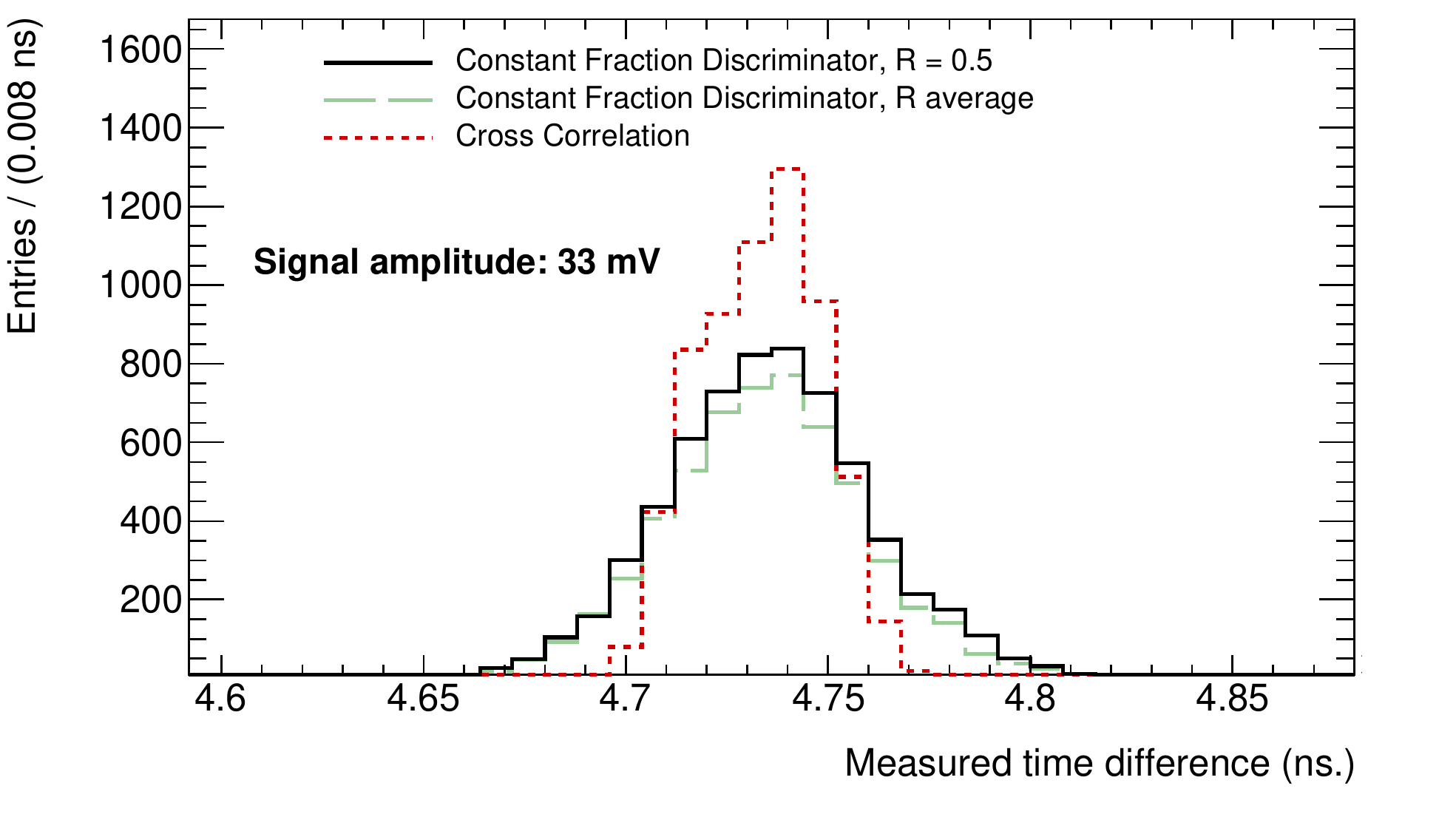}
\includegraphics [width=0.732\textwidth]{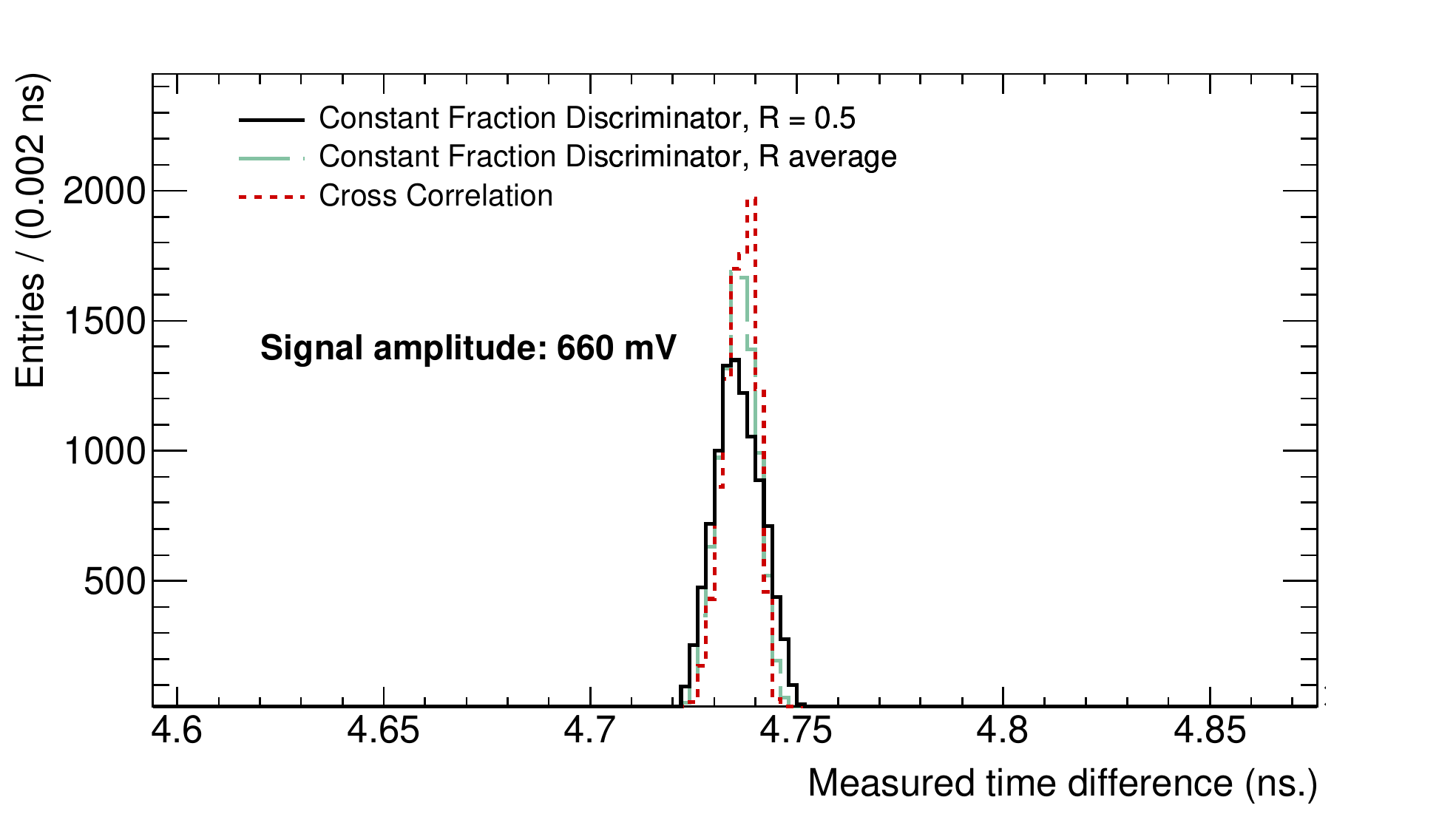}
\caption{Time difference distributions for 10,000 events between two signals of 33 mV (top) and 660~mV (bottom) amplitude delayed by about 5~ns. The results from the three algorithms implemented in the offline software are shown: CFD with fraction $R = 0.5$, refined CFD (labelled as ``$R$ average'') and cross correlation.}
\label{histoPlot}
\end{figure}

\begin{figure}[!htbp]
\centering
\includegraphics [width=0.7\textwidth]{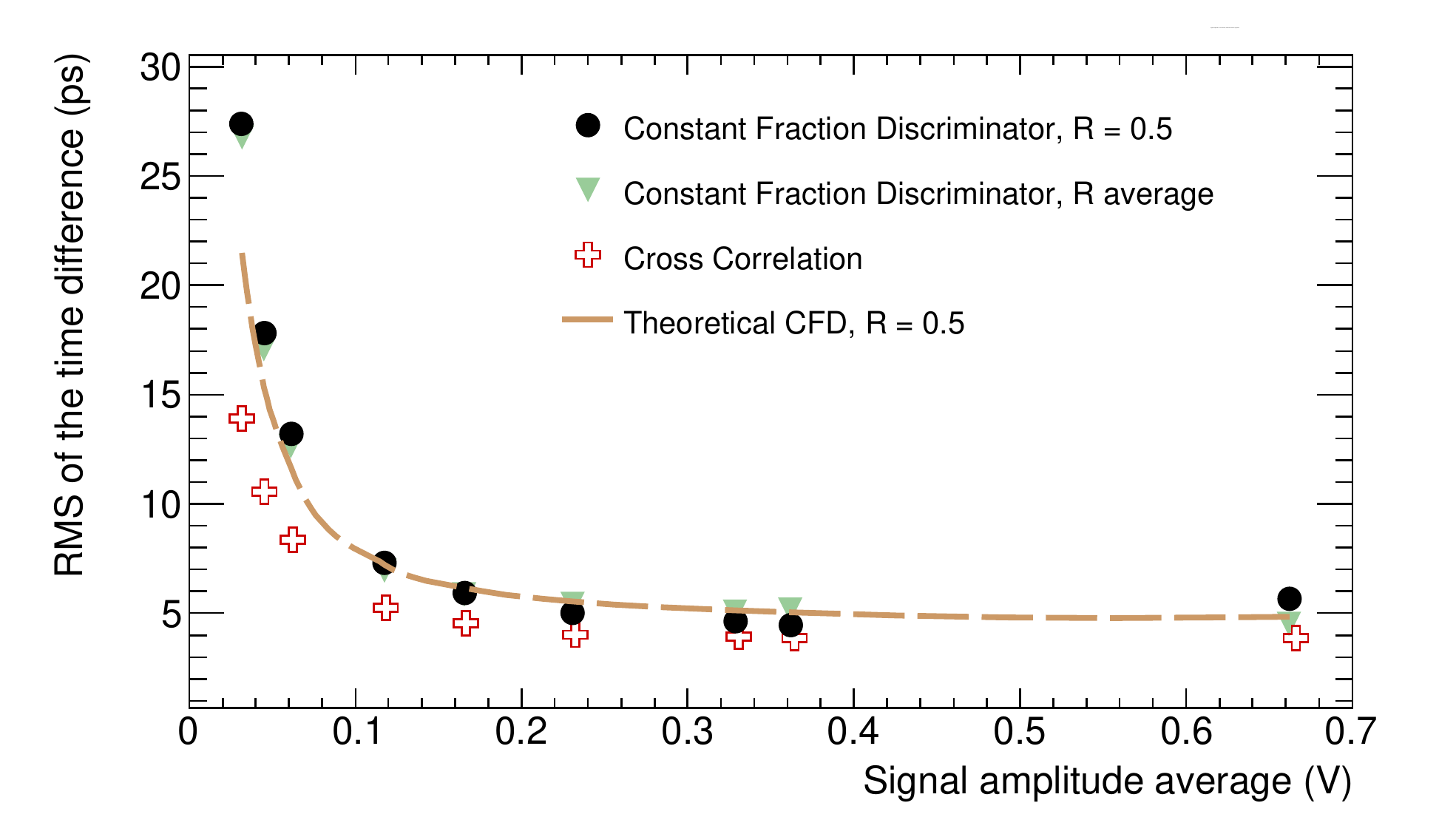}
\caption{RMS of the time difference between two signals measured from 10,000 events by the \textsc{SAMPIC} chip at 6.4 GS/s, as a function of the signal amplitude (amplitude test). The time difference is fixed at about 5 ns. The results from the three algorithms implemented in the offline software are shown: CFD with fraction $R = 0.5$, refined CFD (labelled as ``$R$ average'') and cross correlation.}
\label{ampPlot}
\end{figure}

The histograms of the 10,000 time differences measured with the proposed methods are shown in Figure~\ref{histoPlot} for signals of 33~mV (top) and 660~mV amplitudes (bottom) delayed by about 5~ns (delay test).
The RMS of the time difference distributions obtained for the various amplitudes are shown in Figure~\ref{ampPlot}. At high SNR ($>$~200~mV) the RMS ranges from 4 to 6 picoseconds, the exact numbers varying slightly with respect to the algorithm used and the signal amplitude. However, the RMS increases significantly for amplitudes below 100~mV because of the low SNR. In this region, the cross correlation improves significantly the resolution, up to a factor close to 2. The refined CFD does not provide any significant resolution improvement, most likely because no more than 1 or 2 points are recorded along the rising edge for those very fast signals.

The experimental results for the CFD with a fraction of $R=0.5$ are in good agreement with the theoretical predictions from Equation~\ref{eq:cfd} for those ideal (very fast) signals, except at very low SNR where non-linearities in the rising edge may occur.\footnote{The theoretical predictions are computed for a time difference between two channels, \textit{i.e.} $\Delta t = t_1 - t_2$ with Equation~\ref{eq:cfd} applying for each of the two terms.} The predictions are computed using the observed RMS noise in each channel, which is 1.2~$\pm$~0.1~mV in both cases. The jitter value on the time difference is set to match the predictions to the measurement at high SNR and is found to be about 4~ps RMS. If both channels are uncorrelated, this is equivalent to a $4/\sqrt{2}\simeq 3$~ps resolution per channel. Since small correlations could arise because the same chip is used for both measurements, we only quote the RMS of a $\Delta t$ measurement (\textit{i.e.} 4~ps). This very low value validates the \textsc{SAMPIC} design, which aimed for a jitter below 5~ps RMS (see Section~\ref{sampic}).

\begin{figure}[!htbp]
\centering
\includegraphics [width=0.7\textwidth]{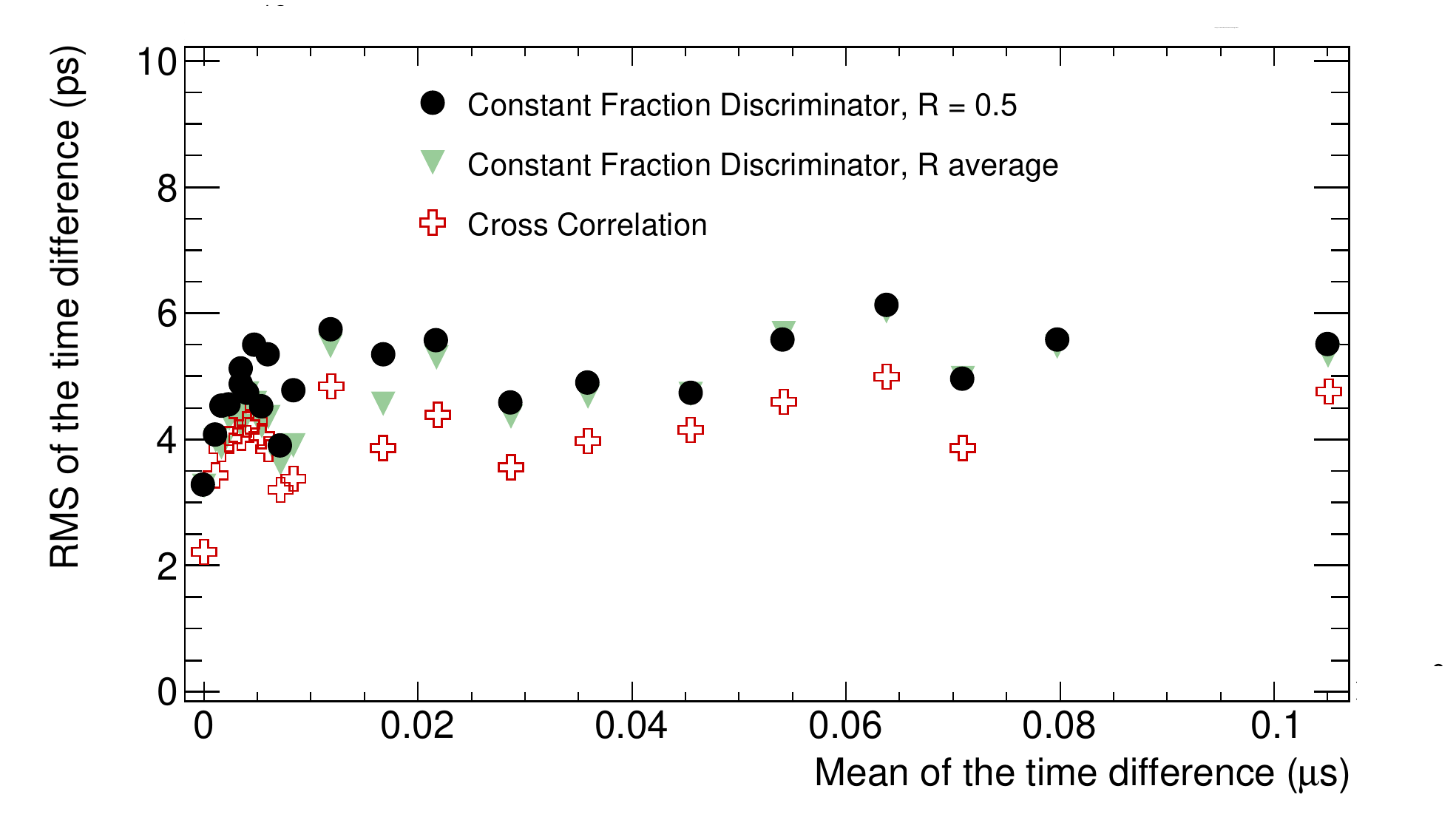}
\caption{RMS of the time difference between two signals measured from 10,000 events by the \textsc{SAMPIC} chip at 6.4 GS/s, as a function of the mean delay (delay test). The signal amplitude is fixed at about 660 mV. The results from the three algorithms implemented in the offline software are shown: CFD with fraction $R = 0.5$, refined CFD (labelled as ``$R$ average'') and cross correlation.}
\label{rms_dt}
\end{figure}

In addition, the RMS has a very small dependence on the delay between the signals (delay test), as shown in Figure~\ref{rms_dt}. 
The small increase of the RMS observed in the Figure is attributed to the attenuation of the signal due to the use of longer cables, as mentioned previously. 

Hence, under those ideal conditions, the \textsc{SAMPIC} chip shows an excellent intrinsic resolution of about 4 ps for signal amplitudes of 0.4~V, in agreement with the design goal. Moreover, the cross correlation algorithm achieves significantly better performance for signals with low SNR with respect to the CFD. In Section~\ref{sec2}, the tests performed using ultra-fast silicon detectors are presented.

\subsection {Tests with ultra-fast silicon detectors}
\label{sec2}

In order to test the SAMPIC chip and the offline software with signals generated by real particle detectors, a particular design of Low Gain Avalanche Diodes-based detectors (LGAD) are used~\cite{CNM} called ultra-fast silicon detectors (UFSDs)~\cite{HFWS, Cartiglia1}. Low Gain Avalanche Diodes (LGAD) are a novel type of silicon detectors that combine the advantages of internal gain, as in avalanche photo-diodes (APDs), with the
properties of standard silicon detectors. The key idea is that the low gain will
not generate additional dark counts and excessive leakage current as it happens in APD,
but will offer an enhanced signal that can be used for timing applications.
Starting from this idea, Ultra-Fast Silicon detectors
were proposed, optimizing the LGAD idea for timing measurements. 


    
   \subsubsection{First setup}

In the first instance, two UFSDs are illuminated by a pulsed infra-red laser. The outgoing signals are then amplified by CIVIDEC\footnote{http://www.cividec.at/} amplifiers.
The use of an infrared laser eases the implementation of the tests, however one has to be aware that high energy particles ($>>$ keV) are expected to induce very different charge carrier fluctuations compared to infrared photons. Therefore, one naively expects a degradation of the timing resolution in the latter case. The expected differences include overall charge fluctuations, which lead to time walk and are in principle corrected by the signal algorithms introduced in Section~\ref{sec:time}, as well as charge density fluctuations, which are at this stage difficult to correct.
The setup is the following :

\begin {itemize}

\item A 1060 nm picosecond laser beam\footnote{PiLas PiL106X.} is split and sent to two UFSDs through optical fibres. The laser has a bandwidth of 2~nm and includes a tunable amplitude. The jitter between the two laser pulses is smaller than 3~ps RMS.

\item Ultra-fast silicon detectors of gain 10~\cite{Cartiglia1,cartiglia:in2p3-01082052} polarized with a voltage of 800~V are used. 

\item The signals are amplified with \textsc{CIVIDEC} C2 broadband amplifiers (BDA). BDAs read the currents generated by the detectors on the amplifier input resistance (50 $\Omega$) and amplify them at 40 dB. The amplifier has a bandwidth of 2 GHz, or, equivalently, a rise time of 180 ps. A C6 fast charge sensitive amplifier (CSA) will be used for the second part of the tests (see Section~\ref{second_setup}).

\item The two active \textsc{SAMPIC} channels are configured to sample data at 6.4 GS/s (C2-BDA) or 3.2 GS/s (C6-CSA).

\item Data is acquired on a computer and processed by the offline analysis software. 

\end{itemize}

\begin{figure}[!htbp]
\centering
\includegraphics [width=0.8\textwidth]{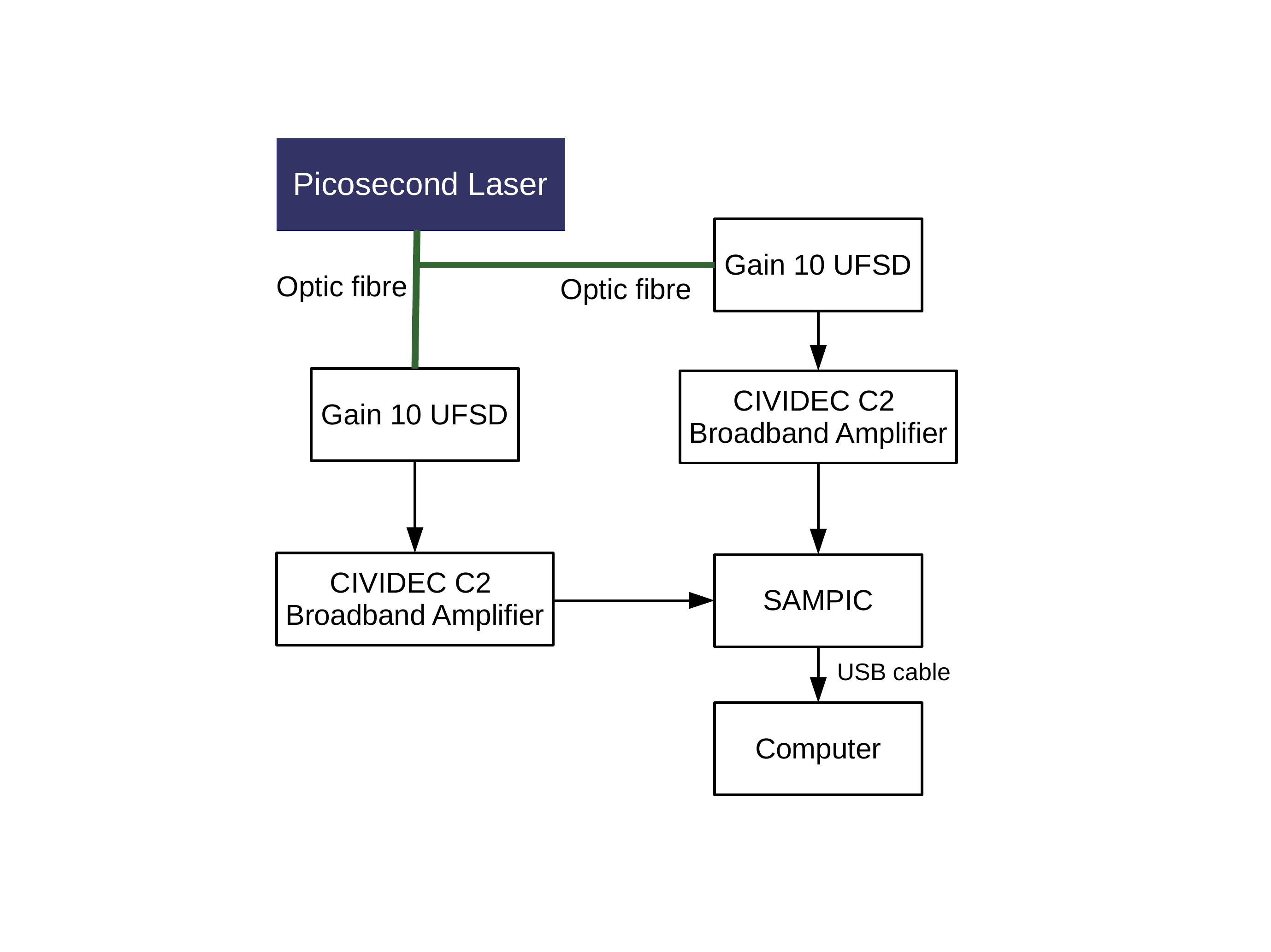}
\caption{Schematic diagram of the first setup (CERN, June 2014)}
\label{setup1}
\end{figure}

A schematic diagram of the first experimental setup is shown in Figure~\ref{setup1}.

\begin{figure}[htbp]
\centering
\includegraphics [width=0.7\textwidth]{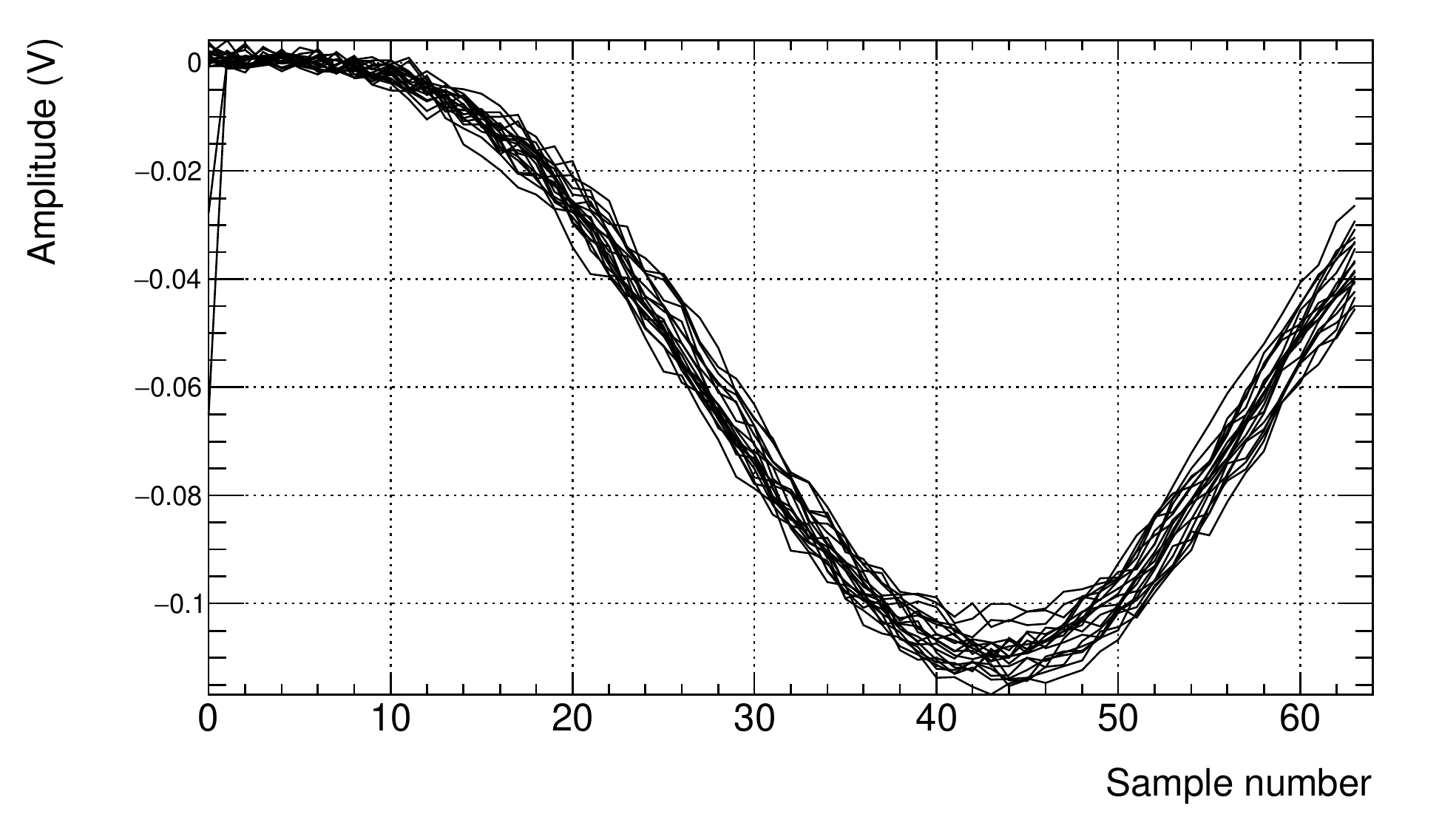}
\caption{Signals generated by a UFSD amplified with a CIVIDEC C2-BDA amplifier and acquired by the \textsc{SAMPIC} chip sampling at 6.4 GS/s. The time spacing between two samples is about 156~ps. Signal properties: 3.5~ns rise time (measured within 10-90\% of the signal amplitude), $\sim 2$ MIPs laser intensity ($\simeq 110$ mV in \textsc{SAMPIC}). The signals are interpolated linearly for plotting purposes. The first point is sometimes lost, which is a known feature of the chip related to the \textsc{SAMPIC} architecture.}
\label{BB_signals}
\end{figure}

\begin{figure}[htbp]
\centering
\includegraphics [width=0.7\textwidth]{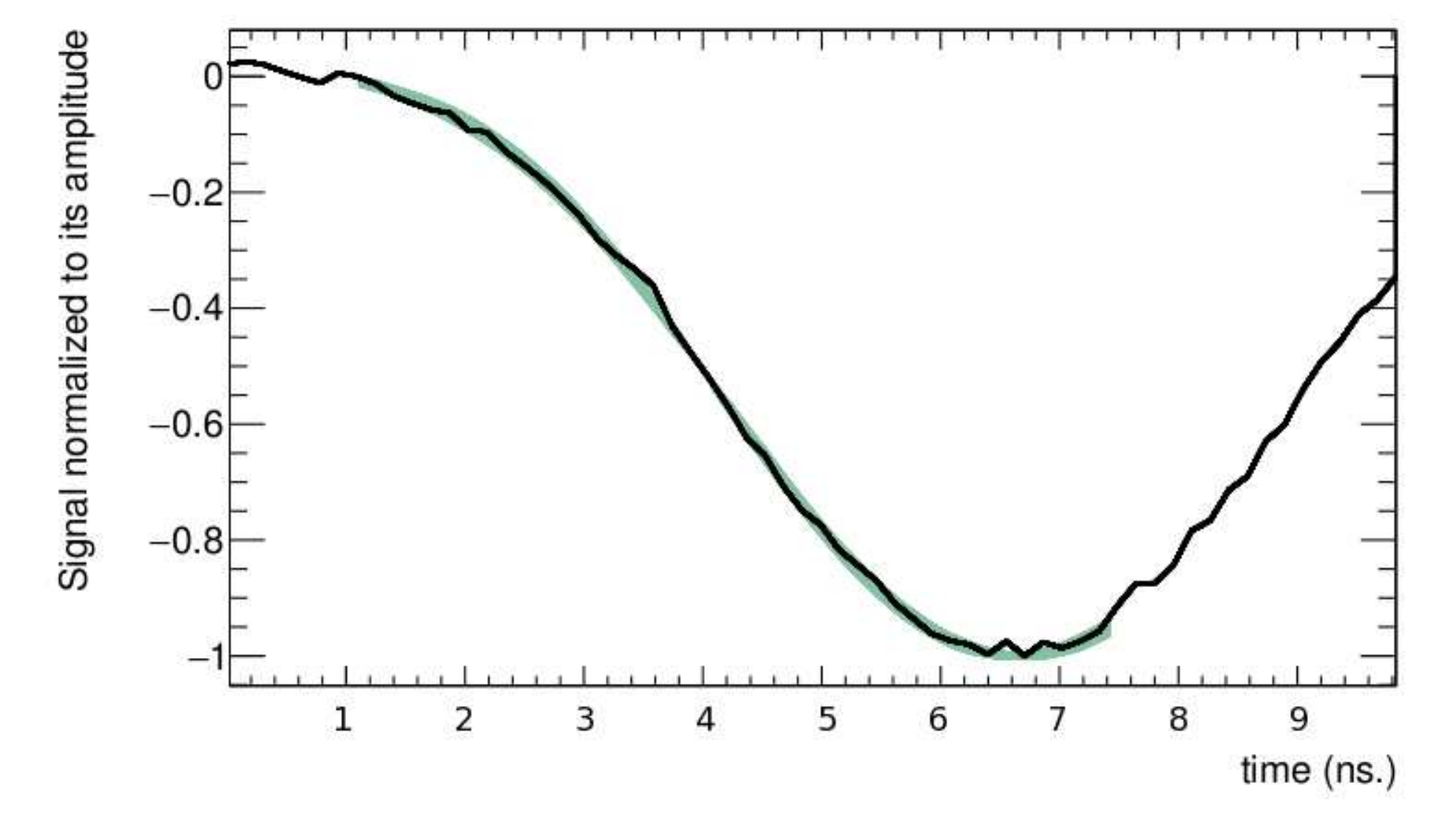}
\caption{Example of synchronization performed between a template (band) and a signal (points) by the cross correlation algorithm. The error band of the template corresponds to its statistical dispersion. The signal corresponds to one of the signals shown in Figure~\ref{BB_signals}.}
\label{sync_CERN}
\end{figure}

Examples of signals acquired with a laser intensity corresponding to the average energy deposited by about 2 Minimum Ionizing Particles (MIPs) are shown in Figure~\ref{BB_signals}. The amplification delivered by the BDAs is not high enough to get a sizeable signal from an energy deposit corresponding to 1 MIP. The polarity of the signal is negative  because the signal is read from the n-side of the UFSD diode. The acquisition window is centred on the rising edge of the pulse. An example of synchronization by cross correlation between a signal and the corresponding template is shown in Figure~\ref{sync_CERN}.

\begin{figure}[htbp]
 \centering
 \includegraphics [width=0.7\textwidth]{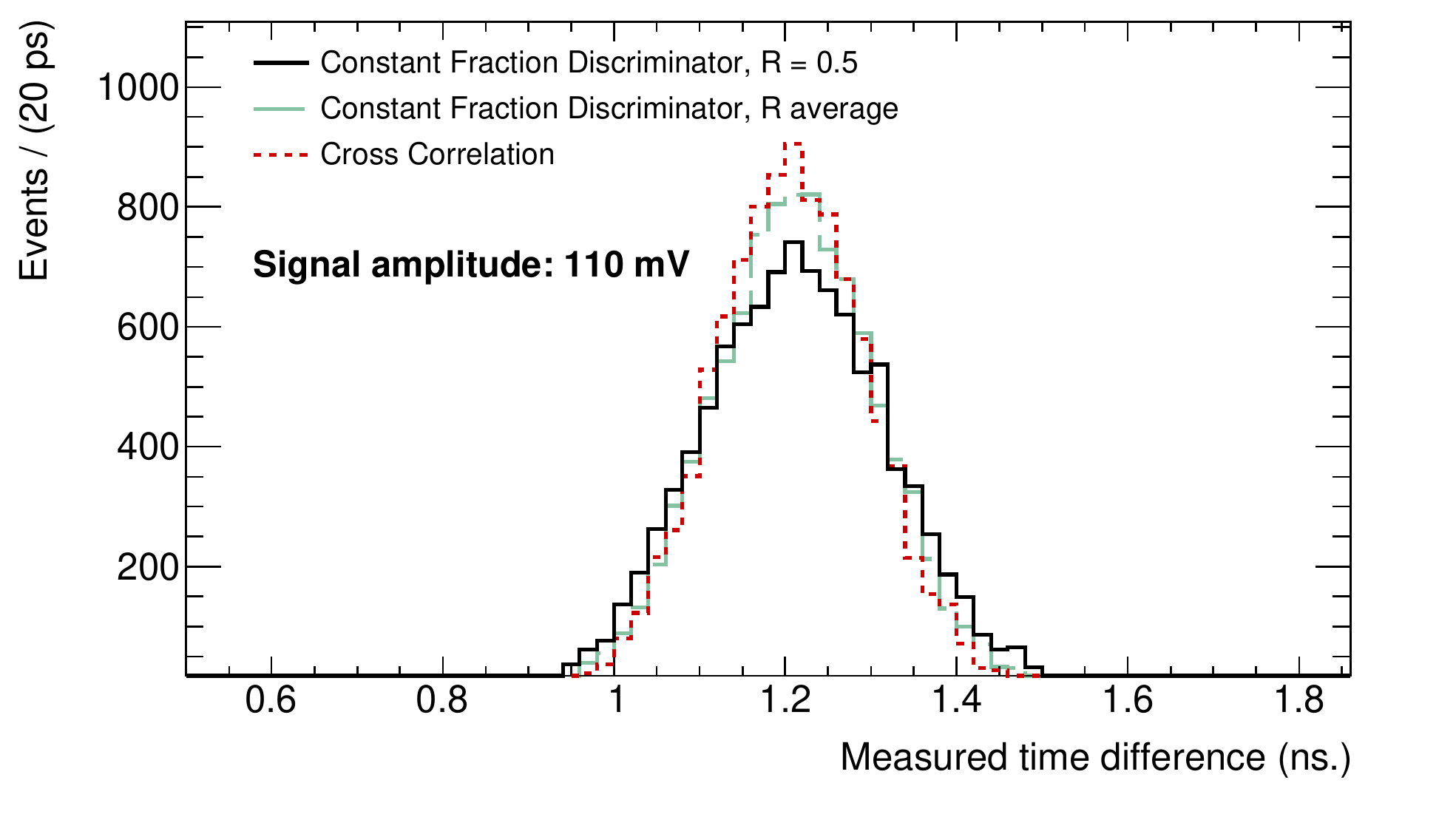}
 \caption{Time difference distributions corresponding to 10,000 signals from C2-BDA-amplified UFSDs of 110~mV of amplitude ($\sim 2$ MIPs). The results from the three algorithms implemented in the offline software are shown: CFD with fraction $R = 0.5$, refined CFD (labelled as ``$R$ average'') and cross correlation.}
 \label{BB_histo}
 \end{figure}

\begin{figure}[htbp]
\centering
\includegraphics [width=0.7\textwidth]{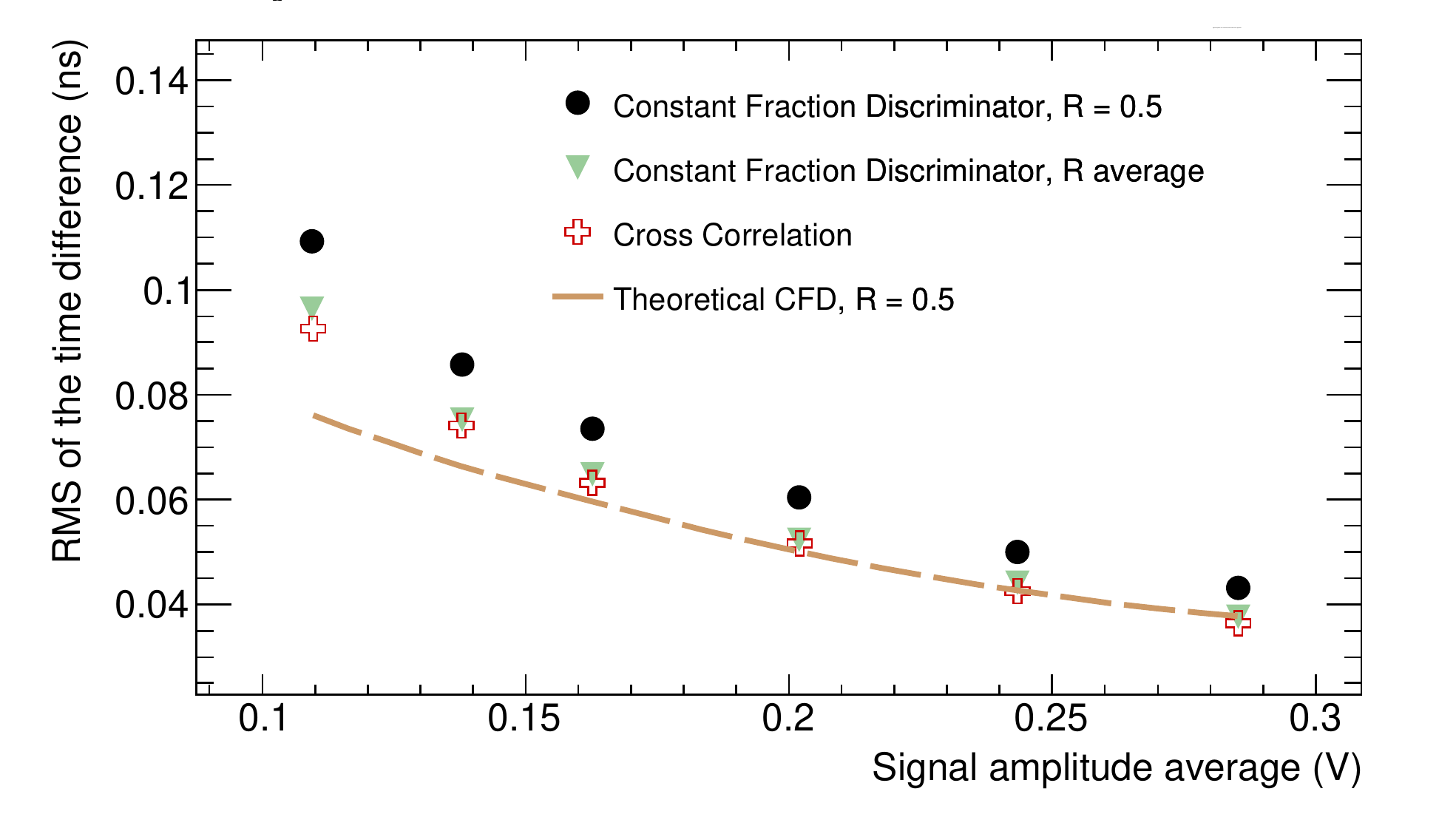}
\caption{RMS of the time difference between two signals from C2-BDA-amplified UFSDs acquired 10,000 times by the \textsc{SAMPIC} chip at 6.4 GS/s as a function of the signal amplitude. Signal amplitudes are varied by tuning the laser intensity to mimic an amplitude corresponding from $\sim 2$ to 4~MIPs. The results from the three algorithms implemented in the offline software are shown: CFD with fraction $R = 0.5$, refined CFD (labelled as ``$R$ average'') and cross correlation.}
\label{ampPlotBB}
\end{figure}

The time difference distributions for 10,000 events according to the various algorithms are shown in Figure~\ref{BB_histo} and the dependence of the RMS of the measured time difference as a function of the amplitude of the signals is shown in Figure~\ref{ampPlotBB}. The measured resolutions are always below 110~ps. 
The RMS obtained from the refined CFD algorithm (labelled as R average) improves the resolution by up to 20~ps with respect to the standard one ($\simeq$~15\%) and gets closer to what one would expect from a standard CFD performed on signals rising linearly (labelled as theoretical CFD). This is most likely due to the fact that much more points are sampled on the rising edge than before, the signals being much slower.

The theoretical expectations are computed from Equation~\ref{eq:cfd} using the RMS noise as observed in data, which is found to be slightly higher than in the previous test ($1.3\pm 0.1$ mV), most likely because of the additional noise coming from the UFSD + BDA. This time, the jitter value is dominated by the detector + amplification system ($\simeq$ 30~ps) and spoils significantly the measurement resolution. The CC algorithm gives results very close to the refined CFD or only slightly better, which is attributed to the fact that the template is defined on a more restricted range around the rising edge compared to the previous test (see Figures~\ref{sync_elec} and~\ref{sync_CERN}). Indeed, the range is chosen in order to exclude from the correlation function the tails of the UFSD signals, which present high statistical fluctuations. Similarly to the previous tests, the resolution reaches a plateau for amplitudes greater than 300 mV.

   \subsubsection{Second setup}
   \label{second_setup}

A second campaign of tests was carried out in October 2014 in the INFN laboratories in Turin. It aimed to measure the timing resolution of a single \textsc{SAMPIC} channel acquiring a signal equivalent to the energy deposit of 1~MIP. In the previous tests, only the resolution on the time difference of signals equivalent to 2~MIPs or more could be accessed. 

\begin{figure}[!htbp]
\centering
\includegraphics [width=0.8\textwidth]{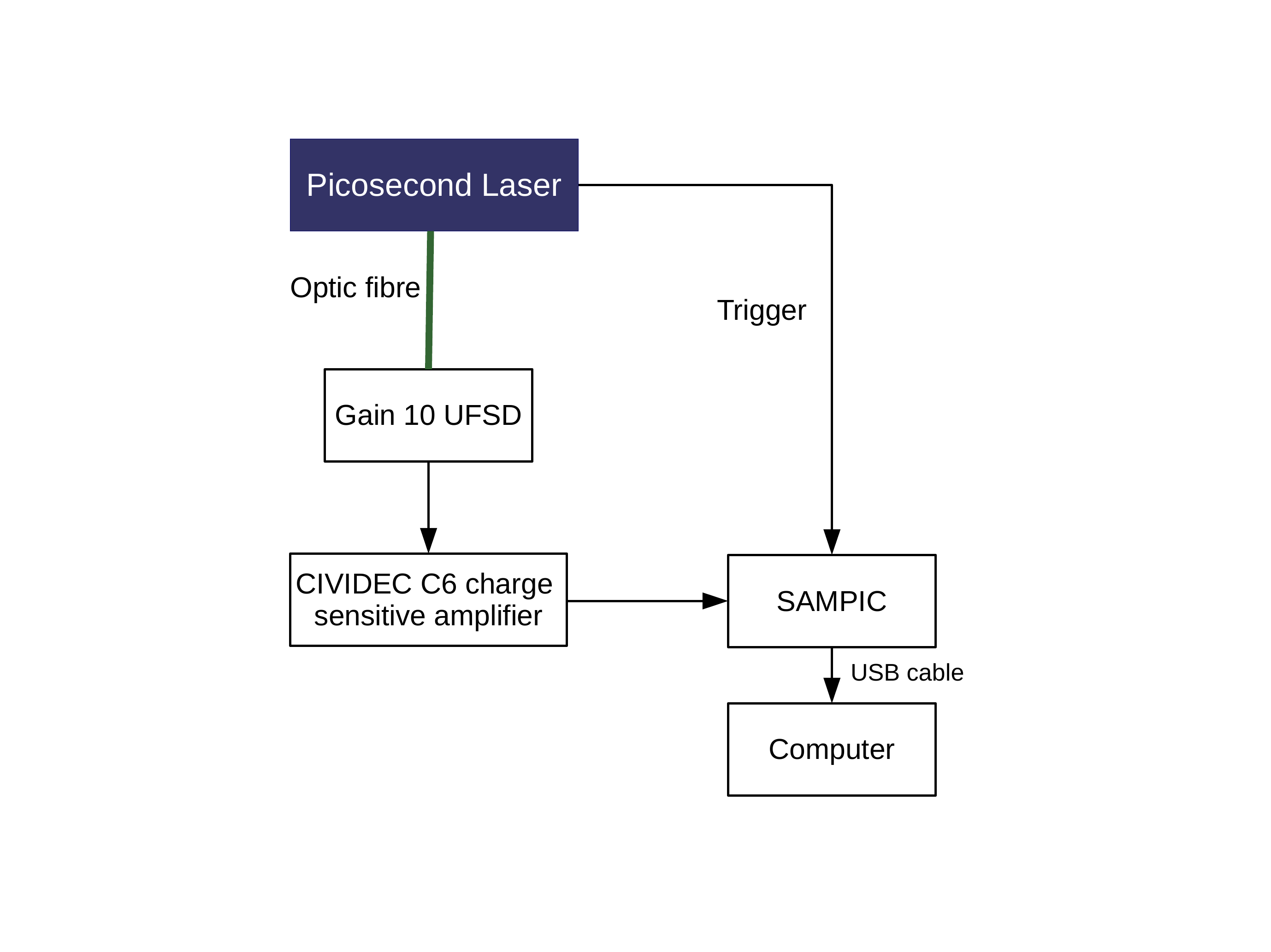}
\caption{Schematic diagram of the second setup (INFN, Turin, November 2014)}
\label{setup2}
\end{figure}

For this purpose, one \textsc{SAMPIC} channel is dedicated to the acquisition of the trigger signal of the laser, whereas the other one acquires a UFSD signal. A schematic diagram of the setup is shown in Figure~\ref{setup2}. Since the trigger signal is very fast and with a high SNR, the time jitter can be considered negligible with respect to the one of the silicon detector. Consequently, the measurements performed in the following evaluates the resolution of a single channel composed of UFSD + CSA + SAMPIC directly.

For this test, the detector is read using a CIVIDEC C6-CSA amplifier so that the laser intensity can be tuned to mimic from 1 to 4 MIP energy deposits. The CSA characteristics depend significantly on the input capacitance. It has a gain of 5.4 mV/fC and a minimum rise time of 3.5~ns. Its SNR is 7.6/fC. 
Signals after amplification range from 120 to 480 mV. Under those conditions, the rise time of the signal sent to SAMPIC is much longer and close to 6~ns (if measured between 10\% and 90\% of maximum). The \textsc{SAMPIC} sampling frequency is therefore decreased to 3.2 GS/s in order to adjust the acquisition time window.

\begin{figure}[htbp]
\centering
\includegraphics [width=0.7\textwidth]{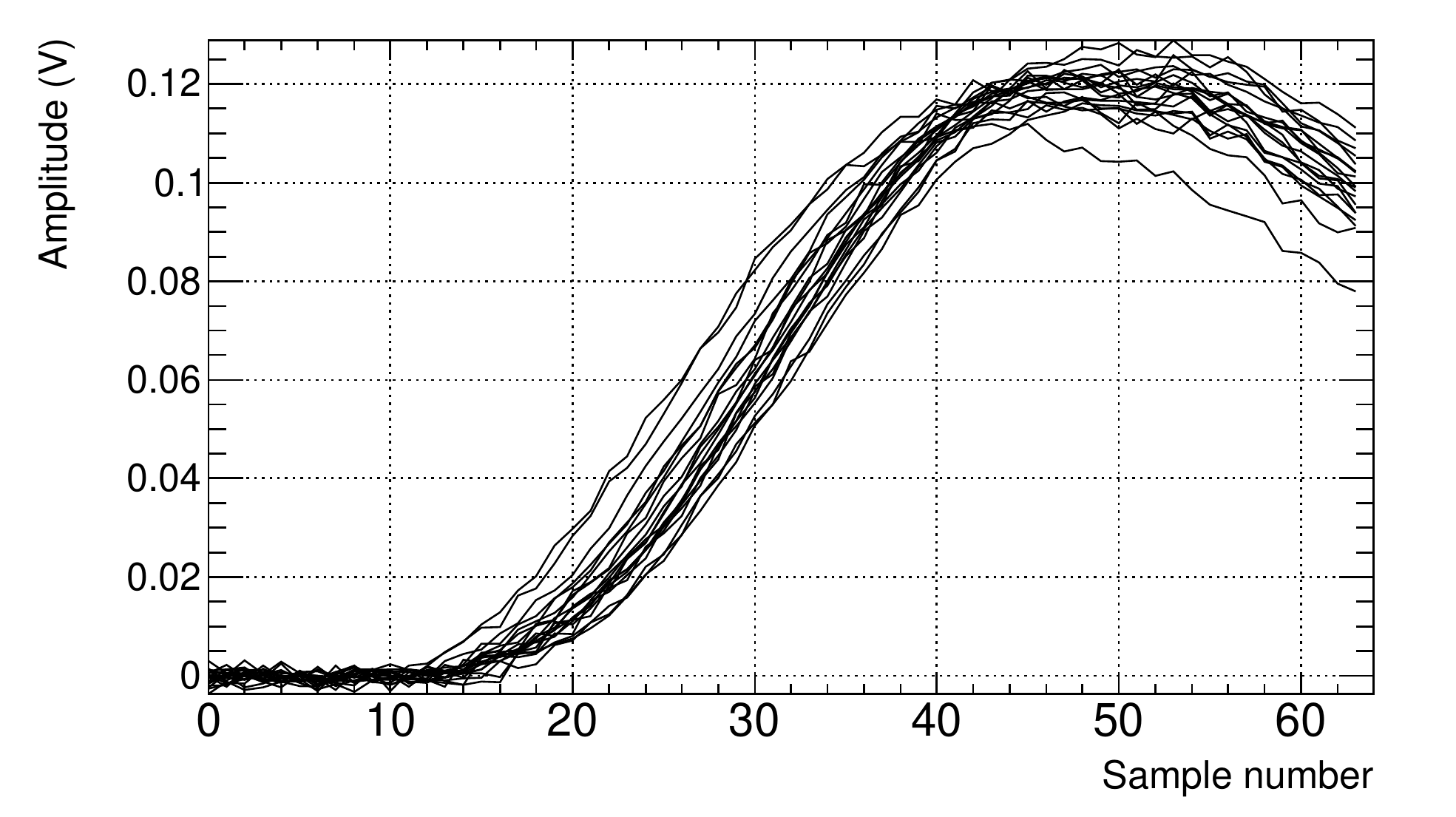}
\caption{Signals generated by a UFSD amplified by a C6-CSA amplifier and acquired by the \textsc{SAMPIC} chip at 3.2 GS/s. The time spacing between two samples is about 312~ps. Signal properties: 6 ns rise time, laser intensity equivalent to 1 MIP ($\simeq$ 120 mV). The signals are interpolated linearly for plotting purposes. The first point is sometimes lost, which is a known feature of the chip related to the SAMPIC architecture.}
\label{CSA_signals}
\end{figure}

\begin{figure}[htbp]
\centering
\includegraphics [width=0.7\textwidth]{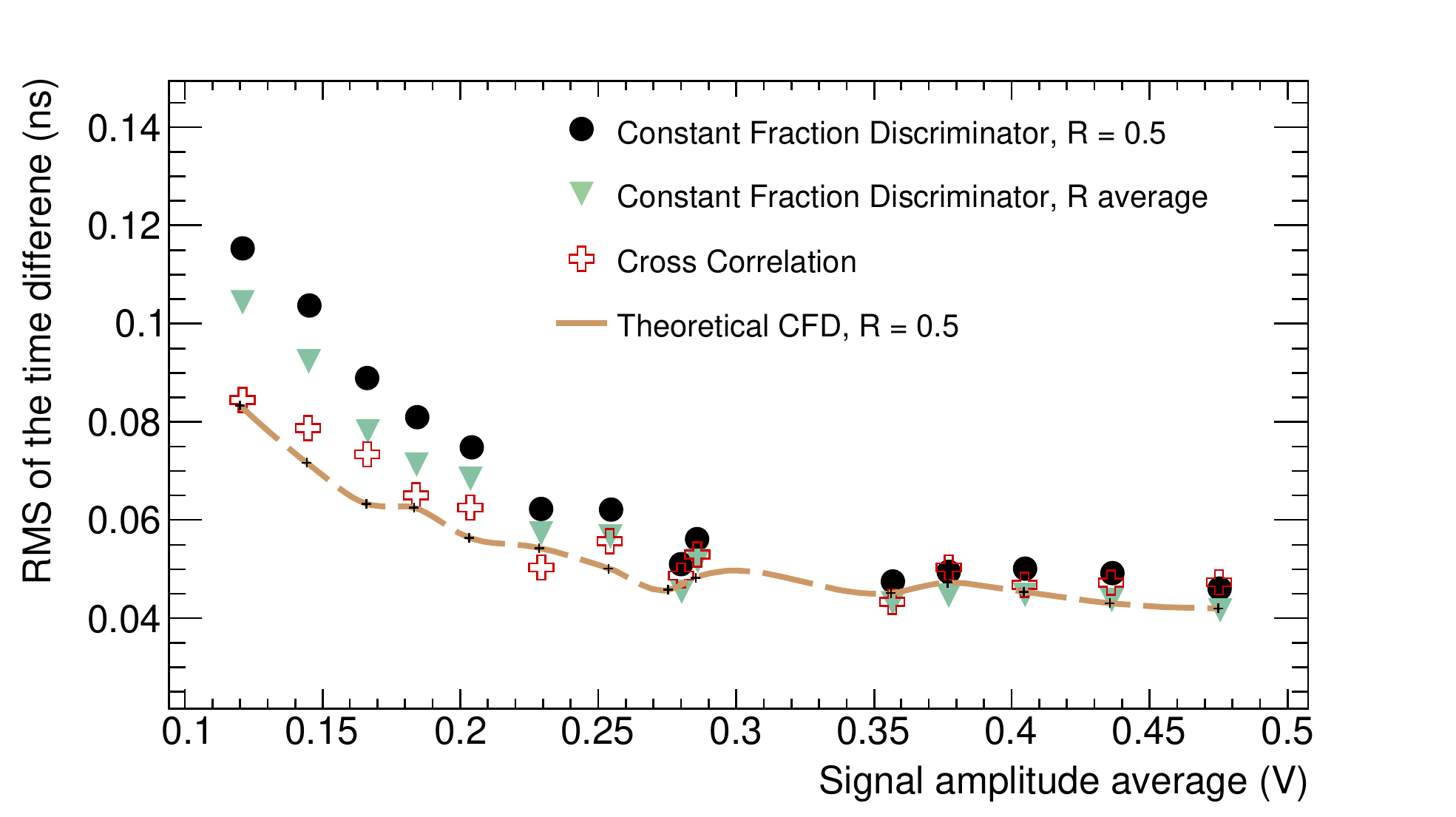}
\caption{Resolution on the timing measurement of a single channel obtained by varying the intensity of the laser pulse. A UFSD amplified by a C6-CSA amplifier is used. 120 mV (480~mV) of amplitude in SAMPIC corresponds approximately to the energy deposit of 1 (4)~MIP(s) with this experimental apparatus. The fluctuations observed for the theoretical CFD and experimental data are due to small fluctuations of the RMS noise and the slow signals. The results from the three algorithms implemented in the offline software are shown: CFD with fraction $R = 0.5$, refined CFD (labelled as ``$R$ average'') and cross correlation.}
\label{ampPlotCSA}
\end{figure}

The signal shape is shown in Figure~\ref{CSA_signals}. 
Similarly to the previous tests, the acquisition window of \textsc{SAMPIC} is centred on the pulse rising edge. The resolution obtained for different signal amplitudes is shown in Figures \ref{ampPlotCSA}. One can see that \textsc{SAMPIC} is able to provide timing measurements up to 85 ps resolution for signals corresponding to 1 MIP, \textit{i.e.} 120~mV amplitude in this case. In this region, the cross correlation algorithm allows to improve the resolution by 20~ps ($\simeq 17\%$) compared to a standard CFD algorithm and by 10~ps compared to the refined CFD. At higher signal-over-noise ratio, \textsc{SAMPIC} can perform measurements with a resolution up to 40 ps whereas the signal rises in 6~ns. As for the previous test, the resolution plateau seems to be reached for amplitudes greater than 300 mV. In the latter case, the different algorithms present similar performance.

The RMS noise and its fluctuations are similar than for the previous test, with values of $1.4\pm 0.1$ mV. However, this time some fluctuations are observed in both the theoretical curve and the experimental measurements. Indeed, the signal being much slower than before, the measurement is also more sensitive to the noise and its fluctuations (see Equation~\ref{eq:cfd}). Again the refined CFD allows to get closer to the theoretical curve whereas the CC does even better at low SNR. The jitter of the system is measured to be close to 40~ps in this case.

In Section~\ref{conclusion}, general conclusions about the tests presented in this paper and plans for future development of the \textsc{SAMPIC} chip are discussed.



\FloatBarrier

\section{Conclusion}
\label{conclusion}

In this article, we reported the first results on the timing measurement resolutions obtained by the \textsc{SAMPIC} V1 chip, which is the version in use as of today. The \textsc{SAMPIC} chip measures the time of the input signals based on their fast sampling. In this paper, experimental results with sampling speed of 6.4 and 3.2 GS/s are reported. In order to analyse \textsc{SAMPIC} data, an online and offline software have been developed to acquire the sampled waveforms and perform accurate timing measurements. In particular, the offline software implements two Constant Fraction Discriminator (CFD) algorithms and one cross-correlation (CC) algorithm which can be compared. Performance studies of various signal processing algorithms applied on waveform data produced by other digitizers than \textsc{SAMPIC} have already been performed, for example in~\cite{Breton2011123}.

The tests were carried out first with the most ideal signals, \textit{i.e.} generated by a very precise signal generator allowing a rise time below 1 ns and a jitter below 1 ps, and then with signals from Ultra-Fast Silicon Detectors (UFSD) illuminated by a pulsed infrared laser. The resolution was found to be largely independent from the delay between the two signals but showed as expected a significant dependence with respect to the signal amplitudes (\textit{i.e.} signal-to-noise ratio). The various results are summarized in Table~\ref{tab:results}.

\begin{table}[!h]
\scriptsize
\centering
    \begin{tabular}{|C{.1\textwidth}|C{.14\textwidth}|C{.14\textwidth}|C{.06\textwidth}|C{.115\textwidth}|C{.115\textwidth}|C{.11\textwidth}|}
      \hline
      & & &  &  &  & \\
       Sampling speed & Type of signal & Measurement & Rise time & CFD resolution (best/worst) & CC resolution (best/worst) & Resolution plateau\\
         & & & (ns) & (ps) & (ps) & (mV) \\
          & & &  &  &  & \\
\hline
 & & &  &  &  & \\
6.4~GS/s & Pulse generator & time difference & $\simeq$ 0.3 & 4/27 & 4/14 & $>$ 200 \\ 
 & & &  &  &  & \\
6.4~GS/s & UFSD ($\simeq$ 2 to 4 MIP)      & time difference & $\simeq$ 3   & 40/95 & 40/93 & $>$ 300 \\
 & & &  &  &  & \\
3.2~GS/s & UFSD ($\simeq$ 1 to 4 MIP)      & single channel  & $\simeq$ 6   & 40/105 & 45/85 & $>$ 300 \\
 & & &  &  &  & \\
    \hline
    \end{tabular}
    \caption{Summary of the results of the tests performed with the \textsc{SAMPIC} chip reported in this paper.}
    \label{tab:results}
\end{table}

At high Signal over Noise Ratio (SNR), all algorithms achieve comparable performance and SAMPIC reaches a resolution of about 4 (40)~ps when using synthesized (UFSD) signals. The difference in resolution observed between synthesized and UFSD signal shows that the resolution in the second case  is limited by the detector and amplification chain and not by the chip. The best \textsc{SAMPIC} resolution performance reaches a value below 5~ps RMS, in agreement with the design goal. For UFSD signals, no significant decrease of  performance was observed when switching from 6.4 to 3.2 sampling frequency given the fairly longer rise time of the signals. The computation time on a 2.3 GHz processor including 4~Gb memory has been found to be about 1~ms (100~ms) per event for the CFD (CC) algorithm. However, this time can be decreased significantly if implemented inside Field-Programmable Gate Arrays (FPGAs), as reported in~\cite{wavecatcher}, where the CFD computation time is only of 10~$\mu$s. The CC algorithm would also highly benefit from parallel computation.

In addition, the cross correlation (which is at first order independent from the baseline computation) and refined CFD algorithms show in general better performance at low SNR. This feature could be very useful to improve the resolution of the measurements without changing the apparatus. It also demonstrates the power of signal processing to improve timing measurement performance and encourages a further development of the offline software.

Finally, the start of the resolution plateau of \textsc{SAMPIC} with respect to the amplitude is close to 300 mV for UFSD signals, which should therefore be the aim in the context of the design of a detection chain willing to use the \textsc{SAMPIC} chip. 

However, one should keep in mind that the experiment apparatus used for the \textsc{SAMPIC} tests reported in this paper differs slightly with respect to the potential applications:

\begin{itemize}

\item For high-energy physics and medical imaging (PET) applications, the expected signal is arising respectively from charged protons and gamma photons, and not infrared photons. Even if the infrared laser intensity is tuned to mimic the average charge deposited by a Minimum Ionizing Particle (MIP) during the tests, the total charge fluctuations as well as the density fluctuations are expected to be different for charged protons. Furthermore, another detector technology than UFSDs is required to detect gamma photons, such as fast scintillating crystals~\cite{Crystals} or detectors based on Cerenkov light~\cite{CALIPSO}, which will behave differently than UFSDs.

\item The minimum dead time per channel for \textsc{SAMPIC V1} is 200~ns, corresponding to an 8-bit conversion. This number will be decreased by a factor of 2 (100~ns) in the next version of the chip thanks to the integration of a Digital-to-Analog Converter (DAC) which will be able to generate Analog-to-Digital Converter (ADC) ramps covering a wider conversion range (7 to 11 bits). The dead time will be further reduced to a value smaller than 5~ns in case of two consecutive hits by using alternatively two SAMPIC channels to digitize one detector channel, similarly to what has been implemented to operate the Analog Ring Sampler (ARS) chip in the ANTARES experiment~\cite{ARS_paper}.

\item The chip dataflow is currently limited by the readout
throughput, whose limit is of about 2~Gbits/s. This corresponds to a full
waveform (64 samples) rate of 2.5~Mevents/s for a full chip. This rate
can be raised above 10~Mevents/s by using regions of interest. Note that
this event rate can be split very unevenly between the channels.
However, in order to use SAMPIC in large scale high-energy physics
experiments, new pre-processing electronics are required in order to reduce
the dataflow generated by the SAMPIC chips in case the
waveform shape is not required by the end user.



\end{itemize}

As a final conclusion, the SAMPIC V1 chip is ready to be used now and has been proved to be a valuable solution for the read-out of low SNR sensors for precise timing measurements. The low power requirements, the cost per channel and the performance of the proposed system are ideal for embedded systems that need the estimation of the time of a sensor event with a precision of tens of picoseconds, including applications not related to high energy physics.

\section{Acknowledgments}
We thank N. Cartiglia for the assistance in using UFSD sensors.
This work has been partially funded by the P2IO LabEx (ANR-10-LABX-0038) in the framework "Investissements d'Avenir" (ANR-11-IDEX-0003-01) managed by the French National Research Agency (ANR).

\clearpage
\bibliography{biblio}

\end{document}